\documentclass[aps,pra,10pt,amsmath,amssymb,
aps,twocolumn,superscriptaddress,showpacs,preprintnumbers]{revtex4-1}

\usepackage{etoolbox}
\usepackage{ulem}

\usepackage{graphicx}
\usepackage{dcolumn}
\usepackage{bm}
\usepackage{color}
\usepackage{soul}
\usepackage{gensymb}

\usepackage{float}

\newcommand{\tj}[6]{ \begin{pmatrix}
  #1 & #2 & #3 \\
  #4 & #5 & #6
\end{pmatrix}}
\DeclareMathOperator\arctanh{arctanh}

\begin{document}

\title{Control of attosecond light polarization in two-color bi-circular fields}

\newcommand{\mbi}{Max-Born Institute for Nonlinear Optics and Short Pulse Spectroscopy, Max-Born-Stra{\ss}e 2A, D-12489 Berlin, Germany}
\newcommand{\icfo}{ICFO - Institut de Ci\`encies Fot\`oniques, The Barcelona Institute of Science and Technology, 08860 Castelldefels (Barcelona), Spain}
\newcommand{\tu}{Technische Universit\"at Berlin, Ernst-Ruska-Geb\"aude, Hardenbergstra{\ss}e 36A, 10623 Berlin, Germany}
\newcommand{\imperial}{Department of Physics, Imperial College London, South Kensington Campus, SW72AZ London, UK}
\newcommand{\humboldt}{Institute f\"ur Physik, Humboldt-Universit\"at zu Berlin, Newtonstra{\ss}e 15, D-12489 Berlin, Germany}

\author{\'A. Jim\'enez-Gal\'an}\email{jimenez@mbi-berlin.de}\affiliation{\mbi}
\author{N. Zhavoronkov}\affiliation{\mbi}
\author{D. Ayuso}\affiliation{\mbi}
\author{F. Morales}\affiliation{\mbi}
\author{S. Patchkovskii}\affiliation{\mbi}
\author{M. Schloz}\affiliation{\mbi}
\author{E. Pisanty}\affiliation{\mbi}\affiliation{\icfo}
\author{O. Smirnova}\affiliation{\mbi}\affiliation{\tu}
\author{M. Ivanov}\affiliation{\mbi}\affiliation{\imperial}\affiliation{\humboldt}

\date{\today}


\begin{abstract}

We develop theoretically and confirm both numerically and experimentally a comprehensive analytical model which
describes the propensity rules in the emission of circularly polarized
high harmonics by systems driven by two-color counter-rotating fields, a fundamental and its second
harmonic.  We  identify and confirm the three propensity rules responsible for the contrast between the
3N+1 and 3N+2 harmonic lines in the HHG spectra of noble gas atoms. We demonstrate
how these rules depend on the laser parameters and how they
can be used in the experiment to shape the polarization properties of the emitted attosecond pulses.

\end{abstract}

\maketitle

\section{Introduction}
Circular or highly elliptical light pulses in the extreme ultraviolet spectral range offer  numerous applications in chiral-sensitive light-matter interactions in both gas and condensed phase, ranging from  chiral recognition \cite{Cireasa2015, Beaulieu2016} to time-resolved magnetization dynamics
and spin currents~\cite{Boeglin2010,Cavalieri2007, Graves2013, Bigot2013, Bigot2009, Stanciu2007, Kirilyuk2010}, not only in
condensed matter but also in isolated atoms \cite{barth2014hole}.
Until recently,
such radiation has only been available at large-scale facilities (e.g. synchrotrons, XFELs), with ultrafast time resolution requiring
free electron lasers.
On the other hand, laboratory-based sources of pulses of attosecond duration ($1$~as $=10^{-18}$~s)
are becoming broadly available,  allowing one
to monitor and control electronic dynamics at their intrinsic (attosecond) timescales \cite{Hentschel2001, Drescher2002, Goulielmakis2010, Sansone2010, Krausz2009, Calegari2014, Gruson2016}. However, the available attosecond pulses are typically linearly polarized, which limits their applicability making them unable
to probe and control such processes as  ultrafast spin and chiral dynamics. These latter require 
the generation of attosecond pulses with controllable degree of polarization.

Ultrashort  XUV/X-Ray pulses are obtained via the high harmonic generation (HHG) process~\cite{Krausz2009},
and their production is linked to the recombination of the electron promoted to
the continuum with the hole it has left in the parent  atom or molecule \cite{Smirnova2009}.
Such recombination is more likely when the electron is driven by a linearly polarized laser field,
leading to linearly polarized harmonics. The suppression of electron return to the parent ion in strongly elliptic
fields means that brute-force approach to the generation of highly elliptically polarized attosecond pulses
by using a highly elliptical  driver fails. One practical solution to this problem is to use
 two-color driving fields in the so-called bicircular configuration, originally proposed
in \cite{Eichmann1995,Zuo1995, Milosevic2000}.
Following elegant recent experiments \cite{Fleischer2014},  this proposal has finally attracted the attention
it has deserved \cite{Fleischer2014,  Pisanty2014, Ivanov2014, Hickstein2015, Medisauskas2015, Kfir2015, Ferre2015, Fan2015,Jimenez-Galan2017, Pisanty2017},
including schemes for the generation of chiral attosecond pulses~\cite{Medisauskas2015,Kfir2016,Hernandez-Garcia2016, Bandrauk2016, Odzak2016, Baykusheva2016}.

\begin{figure}
\centering
\includegraphics[width=\linewidth]{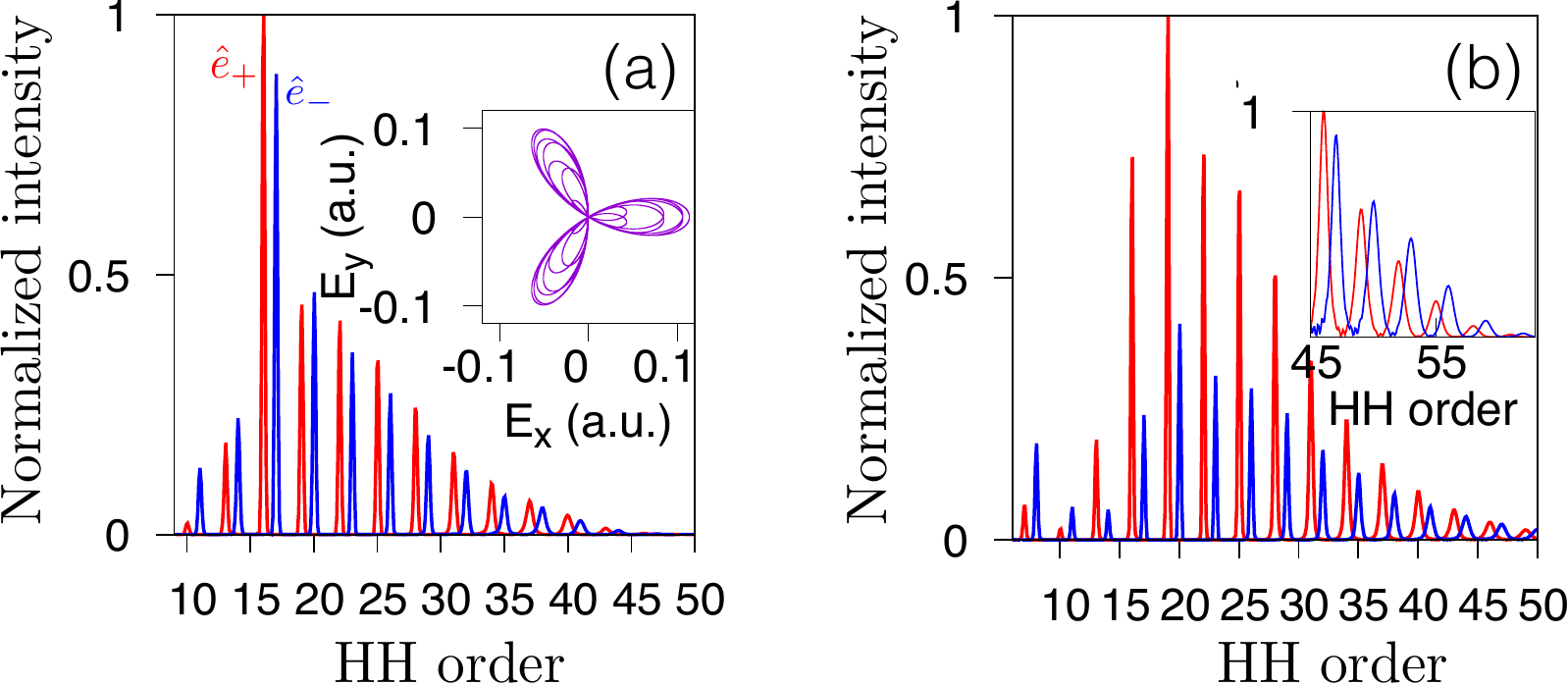}
\caption{Numerical TDSE HHG spectrum of (a) helium and (b) neon.
The inset in (a) shows the y vs x components of the driving electric field. For helium, the fundamental
and second harmonic drivers have field strengths
of $F_{\omega} = F_{2\omega} = 0.056$~a.u., and
duration of 12~fs. For neon, the field strengths of the drivers are
$F_{\omega} = F_{2\omega} = 0.073$~a.u. and
their duration is 12~fs. The red lines indicate the $\hat{\mathbf{e}}_+$ spherical component,
which corresponds to light rotating with the fundamental driver (counter-clockwise), while the blue line indicates the
$\hat{\mathbf{e}}_-$ spherical component, which corresponds to light
rotating with the second harmonic (clockwise). Angular
momentum conservation imposes that the $3N+1$ harmonic lines rotate
with the fundamental field, the $3N+2$ lines rotate with the second
harmonic, and the $3N$ lines are forbidden. The inset in (b) shows the
cut-off harmonics for which the blue lines start to dominate in the spectrum of neon.}
\label{fig:Spectra}
\end{figure}

The bicircular configuration consists of combining a circularly polarized fundamental field
with its counter-rotating second harmonic. The Lissajous figure of the resulting field is shown in the
inset of Fig.~\ref{fig:Spectra}(a). It is symmetric with respect to
a rotation of 120$^{\degree}$. Each of the leaves in the trefoil generates an
attosecond burst, totalling three burst per laser cycle. In the frequency domain, this field produces
circularly-polarized harmonic peaks at (3N+1)$\omega_{IR}$ and (3N+2)$\omega_{IR}$ values,
with the helicity of the fundamental field and the second harmonic, respectively,
while $3N\omega_{\mathrm{IR}}$ harmonic lines are
symmetry-forbidden (see e.g. ~\cite{Milosevic2000, Fleischer2014,  Pisanty2014, Ivanov2014, Hickstein2015, Medisauskas2015, Kfir2015, Jimenez-Galan2017, Alon1998}).

The alternating helicity of  the allowed harmonics is, however, an important obstacle en
route to producing circularly polarized attosecond pulses or pulse trains.
Indeed, to achieve this goal one set of the harmonic lines, e.g. $3N+1$, with well defined helicity,
must  dominate over the other set (e.g. $3N+2$), across a wide range of spectral energies. This is not
always the case, see e.g. \cite{Eichmann1995, Baykusheva2016}  for measurements in argon or helium  or
Fig.~\ref{fig:Spectra}(a) for calculations in helium.
Kfir and collaborators reported substantial suppression of
the (3N+2) harmonic lines of neon by optimizing phase matching conditions in a gas-filled hollow fiber~\cite{Kfir2015,Kfir2016}.
Further analysis demonstrated that such suppression appears already at the microscopic level when the system is
ionized from the 2p orbital of neon  (see Fig.~\ref{fig:Spectra}(b))~\cite{Medisauskas2015,Milosevic2015, Baykusheva2016},
opening the way to practical generation of polarization-controlled attosecond bursts.
Recently, Ref.~\cite{Dorney2017} reported that changing the intensity ratio between the two circular drivers offers a possibility to suppress
even further the $3N+2$ lines.
The exact physical origins of the suppression at the atomic level and the
physical mechanisms responsible for it are, however, not completely understood.
While Ref.~\cite{Medisauskas2015} argued that the origin of the suppression
is linked to the initial angular momentum of the ionizing orbital, Ref.
\cite{Baykusheva2016} suggested the Cooper-like suppression of the
recombination step to be chiefly responsible.

Here, we first present experimental results for neon, showing control over the contrast between the $3N+1$ and $3N+2$ harmonic lines for different relative intensities between the two drivers.
Motivated by these experimental results, we focus on the theoretical investigation of the underlying physical mechanism at the single-atom level, providing a detailed analytical and numerical analysis. We demonstrate
the interplay of three fundamental mechanisms responsible for the different contrast between neighboring harmonic lines.
Their identification offers clear insight into the underlying electronic dynamics
and the possibilities to control the ellipticity of the generated
attosecond pulses.

The first mechanism at play is based on the Fano-Bethe type propensity rules in one-photon 
recombination \cite{Fano1985}.
The second mechanism is traced to the Barth-Smirnova type propensity rules in strong-field ionization
\cite{Barth2011, Barth2013, kaushal2015opportunities}.
The third mechanism is based on the impact of the two driving laser
fields on the continuum-continuum transitions, i.e. the electron dynamics between
ionization and recombination.
The interplay of these three effects links the observed spectral features to the specific
aspects of the sub-cycle electronic dynamics.

In particular, we show that the suppression of the (3N+2) harmonic lines can be controlled
by varying the relative intensity between the two fields, thus providing relatively easy means for
controlling the ellipticity of the attosecond bursts, as suggested in \cite{Dorney2017}.
We confirm our theoretical analysis with the numerical
solution of the time-dependent Schr\"odinger equation (TDSE) and  measurements in neon.

Furthermore, as inferred in \cite{Dorney2017}, we show and explain that the suppression is not restricted to systems with
outer-electrons in $p$ orbitals. Our theory also predicts, and the full solution of the TDSE confirms,
the possibility of significant suppression of the $(3N+2)$ lines in the HHG spectrum of helium,
starting with the ground s-orbital. The suppression is achieved when the intensity of the fundamental field is
sufficiently stronger than that of the second harmonic, eliminating the
misconception that the desired suppression cannot happen when the electron is emitted from a
spherically symmetric orbital, or that it  requires the Cooper-minimum-like suppression of
the recombination matrix elements \cite{Baykusheva2016}.

\section{Experimental observation}

\begin{figure}[htbp]
\centering
\includegraphics[width=\linewidth]{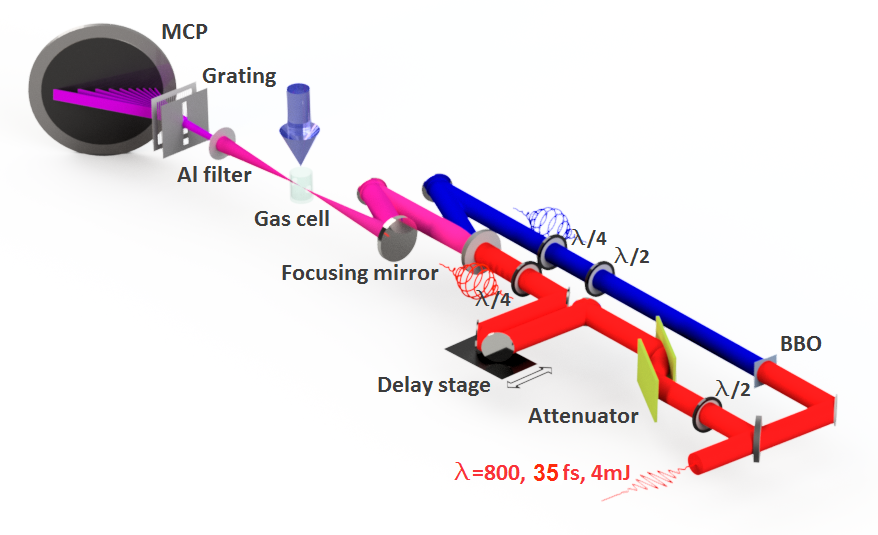}
\caption{Schematic of the optical setup. The BBO (beta-barium-borate) crystal was used 
for the generation of the second harmonic of the fundamental beam, MCP denoted the detector for the XUV radiation,
BS is the beam-splitter.}
\label{fig:setup}
\end{figure}

We begin with the description of our experimental results, which motivate our numerical and analytical analysis. 
To generate the two-colour bi-circular 
fields, we have used a Ti:sapphire-based laser system with a single stage regenerative amplifier
producing 35 fs pulses with up to 4 mJ energy and central wavelength of $\sim795$ nm at 1 kHz repetition rate.
The carrier-envelope phase (CEP) of the pulses was not locked.
The laser beam was directed into the optical setup shown in Fig.~\ref{fig:setup}.
The original beam was split into two beams with 60/40 ratio.
The first, stronger, beam was sent onto a 0.1 mm thin BBO crystal to generate the second harmonic at $\sim400$ nm
with the pulse energy of up to 0.8 $mJ$. The second, weaker, beam remained at a fundamental wavelength of $795$ nm with
the pulse energy up to 1.5 $mJ$. 
The intensity was estimated based on comparing the observed cutoff position with the theoretical calculations
 (which is a standard approach, see e.g. \cite{Smirnova2009}), and taking into account the ratio of pulse energies and focusing conditions for the two incident driving fields.
We could also tune smoothly 
the energies of the two pulses and their ratio by
changing the original input pulse energy and by using a reflective attenuator (Fig.~\ref{fig:setup}).
Both beams were passed through achromatic broadband $\lambda/2$ and $\lambda/4$ waveplates,
yielding nearly circular polarization ($\varepsilon\simeq0.97$) for both the fundamental ($``red"$)
and its counter-rotating second-harmonic ($``blue"$) beams.
The beams were carefully combined in collinear geometry
and focused with a single Ag-mirror at f/100 into a {2-mm}-long gas cell containing target gas.
The cell was initially sealed with a metal foil, which was burned through by the laser beam at the
start of the experiment. The resulting opening $d_0\cong60$ $\mu$m was similar to the focal spot size,
allowing us to keep the gas pressure inside the cell  at $\approx 20-40$ $mbar$ and the vacuum inside the
interaction chamber at the level of $P_{rest}\approx10^{-3}$ $mbar$.
After passing the gas cell, the driving beams were blocked by a $200$ nm thick aluminum foil.
The transmitted XUV radiation was directed towards the XUV spectrometer (for details see \cite{Jimenez-Galan2017}).

\begin{figure}[htbp]
\centering
\includegraphics[width=\linewidth]{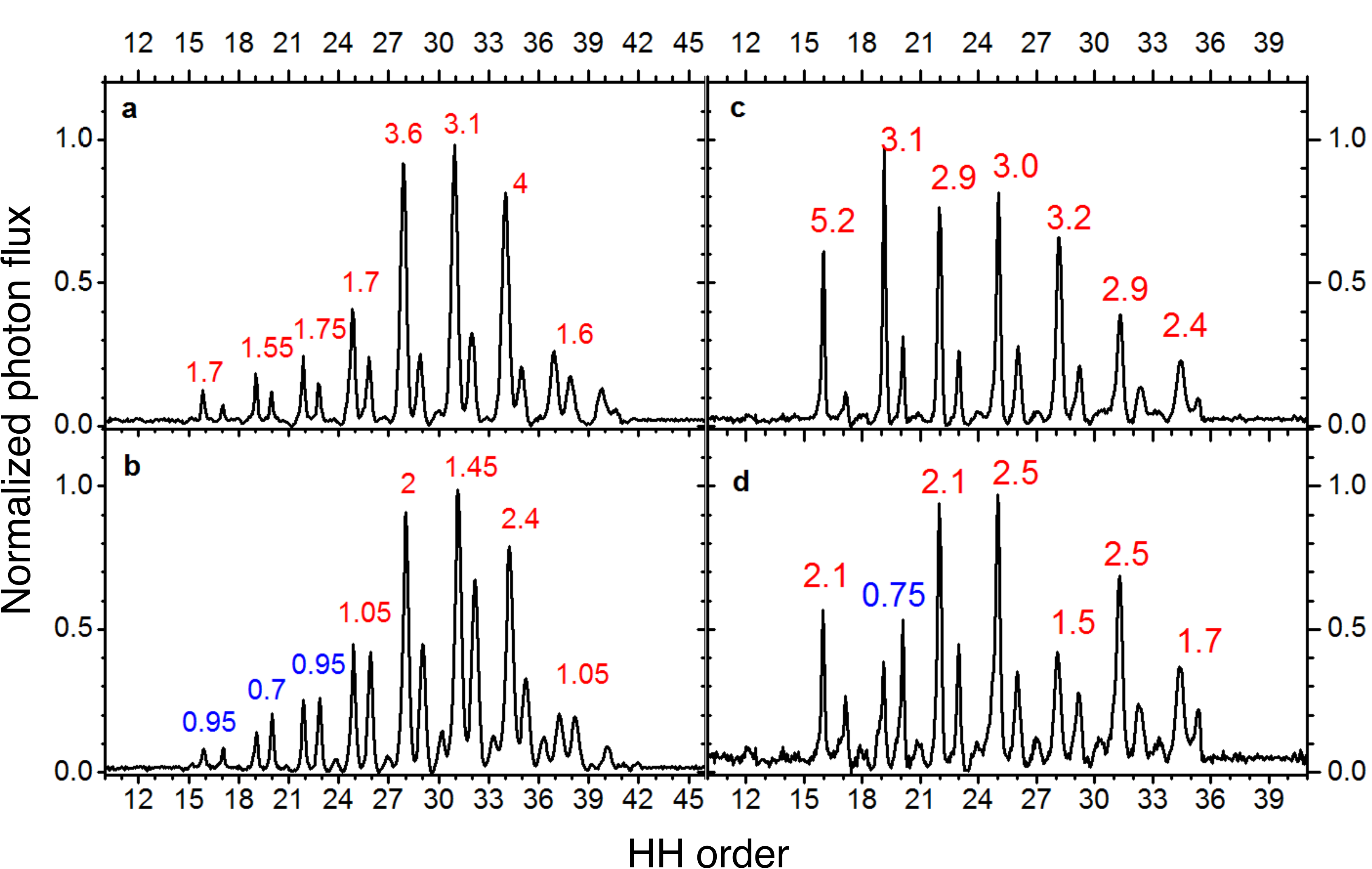}
\caption{High harmonics generated in Neon  for different peak intensities of the fundamental $I_\omega$ and second harmonic $I_{2\omega}$ circularly polarized fields. Intensities (in units of $10^{14}$ $W/cm^2$) are estimated from
the pulse energy, focal diameter, and the cutoff position:
(a) $I_{\omega}\sim 4$, $I_{2\omega}\sim 3.3 $; (b) $I_{\omega}\sim 2.6 $, $I_{2\omega}\sim 3.3 $; (c) $I_{\omega}\sim2.5 $, $I_{2\omega}\sim 2.0$; (d) $I_{\omega}\sim 1.6$, $I_{2\omega}\sim 2.0$.
The ratios of the intensities of the adjacent harmonics with orders $3N+1$ and $3N+2$ are marked in the figures near to the corresponding harmonic pairs.}
\label{fig:NeSp-1}
\end{figure}

The XUV-spectra generated in neon in Fig.~\ref{fig:NeSp-1} show the characteristic structure for the bi-circular scheme.
The harmonics with order $3N$ are strongly suppressed. The allowed $3N+1$ and $3N+2$ harmonics are nearly 
circularly polarised (see results of Ref.~\cite{Fleischer2014, Kfir2015}), rotating in the same direction as the $``red"$ and the $``blue"$ beams, respectively.

\begin{figure}[htbp]
\centering
\includegraphics[width=\linewidth]{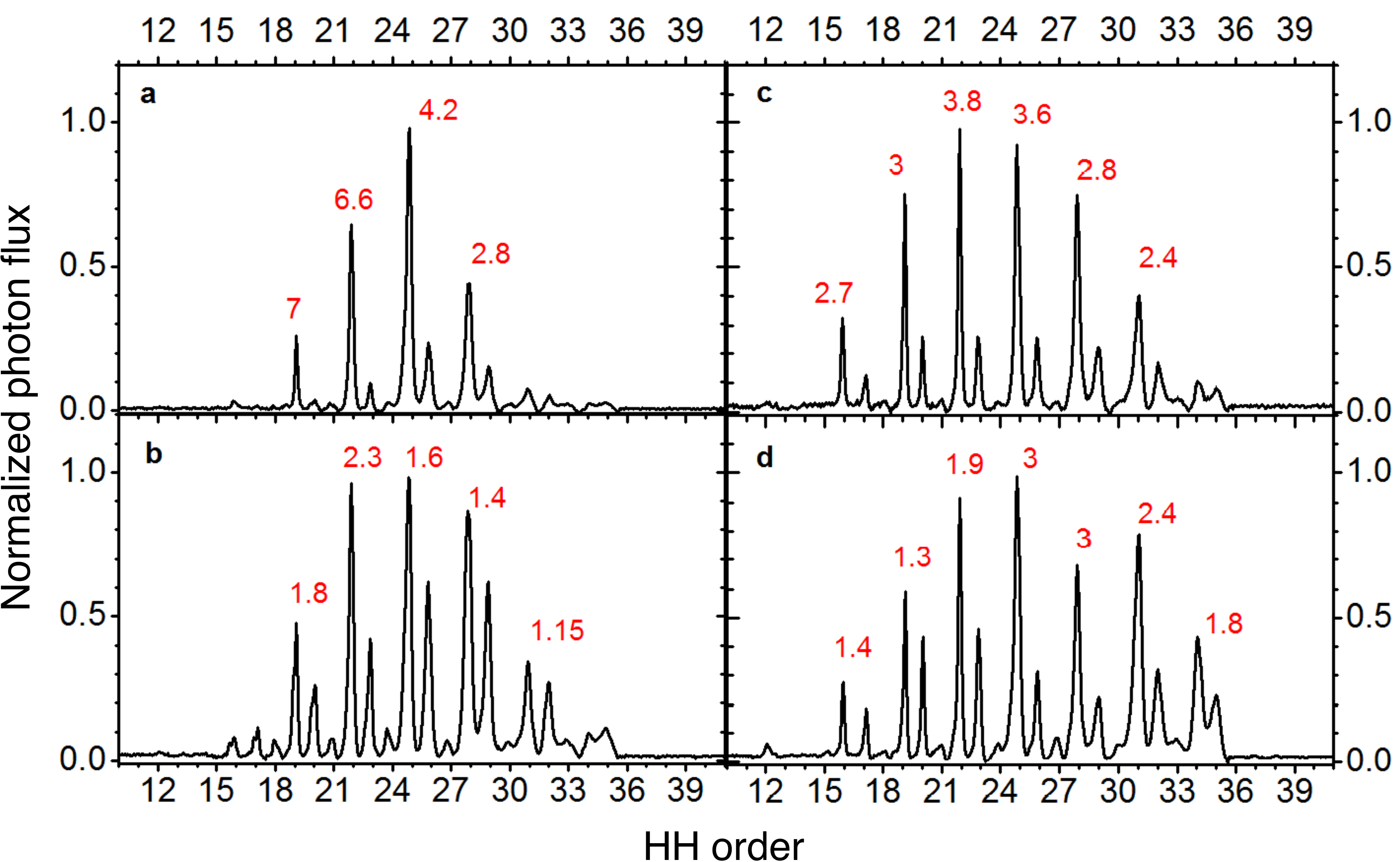}
\caption{High harmonics generated in Neon for different peak intensities of the fundamental $I_\omega$ and 
second harmonic $I_{2\omega}$ fields. Intensities, in units of $10^{14}$ $W/cm^2$, were estimated from
the pulse energy, focal diameter, and the cutoff position:
(a) $I_\omega\sim1.2$, $I_{2\omega}\sim0.7 $; (b) $I_{\omega}\sim 1.2 $, $I_{2\omega}\sim 2.8 $; (c) $I_\omega\sim 1.7 $, 
$I_{2\omega}\sim 0.9 $; (d) $I_{\omega}\sim 1.7 $, $I_{2\omega}\sim 2.0 $.
The ratios of the intensities of the adjacent harmonics  with orders $3N+1$ and $3N+2$ 
are marked in the figure near the corresponding harmonic pairs.}
\label{fig:NeSp-2}
\end{figure}

Figures ~\ref{fig:NeSp-1} and ~\ref{fig:NeSp-2} show spectra of the XUV 
radiation generated for different relative intensities of the fundamental and the second harmonic fields.
The driving fields were changed in a such a way that the intensity of one colour was fixed and the intensity of the 
other was varied.
In Fig.~\ref{fig:NeSp-1} we have varied  the intensity of the fundamental field while in Fig.~\ref{fig:NeSp-2} 
the intensity of the second harmonic field was varied.
In addition, the left and right columns in Figs. ~\ref{fig:NeSp-1} and ~\ref{fig:NeSp-2} present results for different values 
of the fixed fields.

Figures ~\ref{fig:NeSp-1} and ~\ref{fig:NeSp-2}  demonstrate the dependence of the relative intensities
of the $3N+1$ and $3N+2$ harmonics on the ratio between $I_{\omega}$ and $I_{2\omega}$. 
The increase of one driving field intensity relative to the other,
independently of the total field strength, appears to enhance the generation of high harmonics co-rotating 
with it. Therefore, increasing the intensity of, e.g., the red field will change the ratio between the 
neighbouring harmonics and will affect the overall chirality of the generated pulse train.  
Note that if the ratio $I_{\omega}$/$I_{2\omega}$ is kept constant and both intensities are 
changed, the overall harmonic asymmetry is not affected, but the harmonic cut-off scales with the total intensity. 

We also noted that when the intensity of the $``blue"$ beam becomes larger than the intensity of the $``red"$ beam, the 
forbidden $3N$ harmonics become more prominent in the spectrum 
(see figs. ~\ref{fig:NeSp-1} (b) (d) and ~\ref{fig:NeSp-2} (b) (d)).  
We discuss this phenomenon in detail in a separate publication, showing that 
it is related to the breaking of the dynamical symmetry in the system due to excitation of Rydberg states
by the strong blue driver~\cite{Jimenez-Galan2017}.

These experimental results demonstrate a simple practical 
way of controlling  the degree of circularity of the attosecond pulse train by favouring 
harmonics with particular helicity, via changing the intensity ratio of the two driving fields.
Of course, macroscopic effects may play a role in these findings. For example, 
the macroscopic effects can include the contribution of the free electrons to phase matching, 
and the number  of such electrons changes with increasing the intensity of the two driving fields. Nevertheless,
the experimental trends in Figures ~\ref{fig:NeSp-1} and ~\ref{fig:NeSp-2} remain consistent 
across a broad range of intensities of both fields.
Below we show that the dominant role of the $3N+1$ harmonics co-rotating with the red driver 
and the control over the ratio of $3N+1$ vs $3N+2$ harmonics originate 
already at the single-atom (microscopic) level, in accord with the results of 
~\cite{Jimenez-Galan2017, Medisauskas2015}.
We provide detailed analysis of this control and its physical origins using 
a theoretical
model based on the strong field approximation (SFA), and then 
use time-dependent Schr\"odinger equation calculations to further corroborate our 
predictions.

\section{Method} \label{section_method}

We consider a single active electron moving in the binding potential $V_0$ and interacting
with the light field in the dipole approximation. In length gauge,
\begin{equation}\label{eq:hamiltonian}
H= \frac{\hat{\mathbf{p}}^{ 2}}{2} + \hat{\mathbf{r}} \cdot \mathbf{E}(t) + V_0(\hat{\mathbf{r}}),
\end{equation}
where $\hat{\mathbf{p}}$ is the momentum operator, $\hat{\mathbf{r}}$ is the
position operator, and $\mathbf{E}(t)$ is the time-dependent electromagnetic
field for the bicircular configuration, defined by
\begin{equation}\label{eq:electric_field}
\mathbf{E}(t) =\mathrm{Re} \left\{ -F_{\omega} e^{-i\omega t} \hat{\mathbf{e}}_+
+ F_{2\omega} e^{-i 2\omega t} \hat{\mathbf{e}}_{-} \right\}.
\end{equation}
We use atomic units unless otherwise stated. In the above,
$F_{\omega}$ and $F_{2\omega}$ are the electric field strengths
of the fundamental and second harmonic fields, respectively, $\omega$
is the frequency of the fundamental field, and
$\hat{\mathbf{e}}_{\pm} = \mp (\hat{\mathbf{e}}_x \pm i\,\hat{\mathbf{e}}_y)/\sqrt{2}$
are the unit vectors describing the polarization state of the driving fields.
The unit vector $\hat{\mathbf{e}}_+$ corresponds to light rotating in
the counter-clockwise direction, while the unit
vector $\hat{\mathbf{e}}_-$ corresponds to light rotating in the clockwise direction.
In the following, we will use the same notation to indicate the polarization of the generated
high harmonics, and will depict the $\hat{\mathbf{e}}_{\pm}$ component as red/blue lines in the figures.

For the analytical description of the problem, we begin by assuming
the core potential $V_0(\hat{\mathbf{r}})$ in Eq.~\eqref{eq:hamiltonian}
to be short-range, defined by the conditions:
\begin{equation}\label{eq:short_range_potential}
\begin{split}
&H_0 | \psi_{\ell m} \rangle = -I_p | \psi_{\ell m} \rangle, \\
&\langle \mathbf{r} | \psi_{\ell m} \rangle = C_{\kappa \ell} \,\kappa^{3/2}\,(\kappa r)^{-1}\,e^{-\kappa r}\,Y_{\ell m} (\theta_r,\phi_r).
\end{split}
\end{equation}
This approach, based on explicitly using the asymptotic behavior of the
ground state wavefunction
outside the range of the potential (rather than the potential  itself)
is common to many analytical
treatments~\cite{Perelomov1966, Barth2011, Barth2013, frolov2003model, frolov2008effective}, particularly
with the effective range method developed by Frolov, Manakov, and Starace
~\cite{frolov2003model,frolov2008effective} and used extensively to study high harmonic
generation ~\cite{frolov2009analytic, frolov2008wavelength, frolov2011high}.
With this procedure we neglect atom-specific
 features such as Cooper minima or Fano resonances, which can play a role for some atoms in some energy regions~\cite{Baykusheva2016}.
In the above, $\psi_{\ell m}$ is the ground state of the atomic system, with ionization
potential $I_p$, $\kappa = \sqrt{2 I_p}$, $Y_{\ell m}$ is the spherical harmonic of angular and
magnetic quantum number $\ell$ and $m$, respectively, and $C_{\kappa \ell}$ is a dimensionless real
 constant, its exact value is not relevant for our purposes.

The induced temporal dipole, proportional to the response of an individual atom or molecule to the laser field, can be written as
 \cite{Vrakking2014}
\begin{equation}
\label{eq:induced_dipole}
\begin{split}
&\langle \Psi (t) | \hat{\mathbf{r}} | \Psi (t) \rangle =
\sum_{m=-\ell}^{\ell} -i \int_{t_0}^{t} dt' e^{i I_p (t'-t)} \times \\
&\times \int d\mathbf{p}\,e^{-i S_V (\mathbf{p},t,t')} \mathbf{d}_{\ell m}^{({\text {rec}})} [\mathbf{p} + \mathbf{A}(t)] \Upsilon_{\ell m} [\mathbf{p} + \mathbf{A}(t')] + {\rm c.c.},
\end{split}
\end{equation}
where `c.c.' denotes the complex conjugate of the preceding expression,
the sum over $m$ takes into account the $m$-degeneracy of the ground state,
$ \mathbf{p}$ is the canonical momentum,
and $\mathbf{A} (t) = -\int dt\,\mathbf{E}(t)$ is the vector potential of the field
given by Eq.~\eqref{eq:electric_field}.

The Volkov phase $S_V (\mathbf{p},t,t')$ is
\begin{equation}
S_V (\mathbf{p},t,t') = \frac{1}{2} \int_{t'}^t d\tau\,[\mathbf{p} + \mathbf{A}(\tau)]^2,
\end{equation}
and the (scalar) ionization factor is
\begin{equation}
\Upsilon_{\ell m} [\mathbf{p} + \mathbf{A} (t') ] = \left[ \frac{[\mathbf{p} + \mathbf{A}(t')]^2}{2} + I_p \right] \langle \mathbf{p} + \mathbf{A}(t') | \psi_{\ell m} \rangle.
\end{equation}
For the plane-wave continuum, the plane waves are defined as
\begin{equation}
\langle \mathbf{r} | \mathbf{p} + \mathbf{A}(t) \rangle = \frac{e^{i \mathbf{v} (t) \cdot \mathbf{r}}}{(2\pi)^{3/2}},
\end{equation}
where $\mathbf{v} (t) = \mathbf{p} + \mathbf{A}(t)$ is the kinetic momentum.
The recombination dipoles are computed for the instantaneous electron
velocity at the moment of emission $t$,
\begin{equation}
\mathbf{d}_{\ell m}^{({\text{rec}})} [\mathbf{p} + \mathbf{A}(t)]  = \langle \psi_{\ell m} | \hat{\mathbf{r}} |  \mathbf{p} + \mathbf{A}(t) \rangle.
\end{equation}
We note that this theoretical approach  is well adapted to incorporate
exact recombination dipoles (see e.g. \cite{frolov2008wavelength,frolov2011high}).
This expression for the induced dipole is a direct extension of the Perelomov, Popov, and Terent'ev approach
to high harmonic generation ~\cite{Vrakking2014}. The expresssion includes only the processes where the electron recombines
to the same orbital from which it was ionized and assumes that there is no permanent dipole in the ground state.

The HHG spectrum at the frequency $N\omega$ is proportional to the induced frequency dipole,
$\mathbf{D}(N\omega)$,
\begin{equation}\label{eq:def_frequency_dipole}
\begin{split}
&I (N\omega) \propto (N\omega)^4 |\mathbf{D} (N\omega)|^2,\\
&\mathbf{D}(N\omega) = \int dt \langle \Psi (t) | \hat{\mathbf{r}} | \Psi (t) \rangle  \,e^{i N\omega t}.
\end{split}
\end{equation}

The five integrals in Eq.~\eqref{eq:def_frequency_dipole} can be computed using the saddle point method
(see e.g. ~\cite{Vrakking2014}). Amongst all possible electron trajectories,
the saddle point method selects
those that contribute the most to the five-fold integral in
Eq.~\eqref{eq:def_frequency_dipole}, termed `quantum trajectories' (see e.g. \cite{salieres2001feynman}) and reflecting purely quantum under-the-barrier motion~\cite{Lewenstein1994}
of the otherwise classical trajectories, with
complex times of ionization and recombination.
This leads to complex ionization and recombination velocities and to complex ionization and recombination angles, which would be crucial to
the analysis of the high harmonic spectrum and, especially, to the contrast of the
adjacent harmonic lines with opposite helicity.

Applying the saddle point method, we can express the induced frequency dipole as
\begin{equation}\label{eq:FrequencyDipole}
\begin{split}
&\mathbf{D}(N\omega) = \sum_j^{n_s} \mathbf{D}^{(j)} (N\omega), \\
&\mathbf{D}^{(j)}(N\omega) \approx \sum_{m=-\ell}^{\ell} \mathbf{d}_{\ell m} ^{\text{ (rec)}} [\mathbf{p}_s^{(j)} +  \mathbf{A}(t_r^{(j)})] \times \\
& \times e^{-i S(\mathbf{p}_s^{(j)},t_r^{(j)},t_i^{(j)})} \Upsilon_{\ell m} [\mathbf{p}_s^{(j)} + \mathbf{A} (t_i^{(j)})] e^{i N\omega t_r^{(j)}}.
\end{split}
\end{equation}
The quantities $\mathbf{p}_s, t_r$ and $t_i$ are the complex-valued saddle point solutions for the momentum, and times of recombination and ionization, respectively.
The sum of index $j$ runs over all the $n_s$ stationary points. We will consider
only those stationary points that correspond to the so-called short trajectories,
which we can easily identify \cite{Milosevic2000}.
In this case, $n_s = 3M$, where $M$ is the number of laser cycles.

Let us consider the contribution of
a single laser cycle. The spectral intensity is the coherent sum of the
three consecutive bursts. Since we are interested in the contrast between the two helicities
of the emitted light ($\hat{\mathbf{e}}_+$ and $\hat{\mathbf{e}}_-$) in the spectra, it is useful to
separate the intensity of the emitted light into the two helical components,
\begin{equation}\label{eq:SpectrumIntensity}
I(N\omega) = I_{+}(N\omega) + I_{-} (N\omega), \\
\end{equation}
where
\begin{equation}
I_\pm (N\omega) = \left | A_\pm^{(1)} e^{i\phi_{\pm}^{(1)}} +  A_{\pm}^{(2)} e^{i\phi_\pm^{(2)}} + A_\pm^{(3)} e^{i\phi_\pm^{(3)}} \right |^2,
\end{equation}
with the amplitude and phase of the bursts corresponding to the amplitude and
the phase of the frequency dipole in Eq.~\eqref{eq:FrequencyDipole},
\begin{equation}\label{eq:SpectrumIntensity_Definitions}
A_{\pm}^{(j)} = | D_{\pm}^{(j)}(N\omega)  | \quad \text{and} \quad \phi_\pm^{(j)} = \arg \left[D_{\pm}^{(j)}(N\omega) \right].
\end{equation}
Above, $D_{\pm}$ is the $\hat{\mathbf{e}}_{\pm}$ helical component of the dipole vector $\mathbf{D}$, i.e., $D_{\pm} = \mathbf{D} \cdot \hat{\mathbf{e}}^*_\pm$, with $\hat{\mathbf{e}}_{\pm} \cdot \hat{\mathbf{e}}^*_{\pm}=1$.
In the long pulse limit, we have $A_\pm^{(1)} = A_\pm^{(2)} = A_\pm^{(3)} \equiv A_\pm$, and $\phi_\pm^{(j)} - \phi_\pm^{(j-1)} =  (2\pi/3) (N \mp 1)$ (see Appendix \ref{app:amp&phase}), so
that the intensity can be written as
\begin{equation}\label{eq:intensity}
I_\pm (N\omega) =  A_\pm ^2 \left | 1 + 2\cos \left[\frac{2\pi}{3} (N \mp 1)\right] \right|^2.
\end{equation}
The second term on the right-hand side of \eqref{eq:intensity} is responsible for
the well-known selection rules, i.e., harmonics $3N\pm1$ are favored for $\hat{\mathbf{e}}_\pm$ and $3N$ harmonics are suppressed. The relative strengths of the harmonic lines,
however, and thus the contrast between the $(3N+1)$ and $(3N-1)$ lines is
contained exclusively in the amplitude $A_\pm$.

According to Eq.~\eqref{eq:FrequencyDipole}, we may write,
\begin{equation}\label{eq:dipole_reduced}
\begin{split}
&A_{\pm}^2 = e^{2(\Im \{S\} - N\omega \Im\{t_r\})} \times \\
&\times \left| \sum_{m=-\ell}^{\ell} \left(\text{d}_{\ell m, \pm} ^{({\rm rec})} [\mathbf{p}_s +  \mathbf{A}(t_r)] \right)\,\Upsilon_{\ell m} [\mathbf{p}_s + \mathbf{A} (t_i)] \right|^2,
\end{split}
\end{equation}
where $\Im\{x\}$ denotes the imaginary part of $x$ and we have assumed that both the imaginary return time and the action do not change considerably from burst to burst, i.e.,  $\Im \{t_r^{(1)}\} \approx \Im \{t_r^{(2)}\} \approx \Im \{t_r^{(3)}\} \approx \Im \{t_r\}$ and $\Im \{S^{(1)}\} \approx \Im \{S^{(2)}\} \approx \Im \{S^{(3)}\} \approx \Im \{S\}$. Since  $\exp\left[2(\Im \{S\} - N\omega \Im\{t_r\}\right]$ is a scalar independent of $m$, it does not influence the contrast
between  the two helical components in the spectrum and we can ignore it
for our purposes.

\section{Propensity rules in two-color HHG}

In this section we will analyze how and why the strength of harmonic lines varies across the HHG spectrum of atoms emitting $s$-shell and $p$-shell electrons in the bicircular scheme.
In particular, we will concentrate on two systems: helium and neon, whose HHG spectrum
obtained by solving numerically the TDSE is
shown in Fig.~\ref{fig:Spectra}. 

To solve the TDSE we used the code described in \cite{Patchkovskii2016}. To simulate the neon atom, we used the 3D single-active electron pseudo-potential given in \cite{Tong2005}, with a value of $z_{eff} = 10$
at the point $r=0$. To eliminate the contribution of long trajectories and bound states, and save computational time, we have used a small radial box of 70~a.u. The total number of points was $nr=2500$, we used a uniform grid spacing of 0.02~a.u. and a complex boundary absorber at $50$~a.u. The time grid had a spacing of $dt = 0.0025$~a.u. and the maximum angular momenta included in the expansion was $\ell_{\text{max}} = 60$. All the discretization parameters have been checked for convergence.

Eq.~\eqref{eq:dipole_reduced} has now reduced our problem to the study of the ionization and recombination matrix elements within a single burst. Let us focus first
on the one-photon recombination matrix element. To begin with, we
point out one key aspect, which distinguishes
one photon recombination in high harmonic emission from standard
one-photon ionization and/or recombination process.
This key aspect is often overlooked and implicitly ignored
when one states that the recombination step in  HHG is
the (time-reversed) analog of one-photon ionization. The difference, however,
is clear: in HHG recombination is conditioned on the electron return to
the parent core and therefore carries in it the imprint of the ionization step.
In the quantum trajectory analysis, this imprint is encoded into
the complex-valued recombination time $t_r$ and (in general)
the complex-valued velocity ${\bf v}(t_r)$. Below we will denote the real and
imaginary parts of these and associated quantities with one or two `primes'
correspondingly, e.g. $\Re\{t_r\}=t_r', \Im\{t_r\}=t_r''$, etc.

For the one-photon dipole
transition we apply the
Wigner-Eckhart theorem and separate the radial part from the angular part:
\begin{equation}\label{eq:recombination_matrix_element}
\begin{split}
&\text{d}_{\ell m, \pm}^{(\rm rec)}[\mathbf{p}_s +  \mathbf{A}(t_r)] \equiv \langle \psi_{\ell m} | \hat{\mathbf{d}} \cdot \hat{\mathbf{e}}^*_\pm | \mathbf{v} (t_r) \rangle = \\
& = \sum_{\ell'=0}^{\infty}\,\mathcal{R}_{v(t_r), \ell'} e^{-i (m\pm1) \phi_{v(t_r)}'}  \tj{\ell}{\ell'}{1}{-m}{(m\pm 1)}{\mp 1}  \times \\
&\times \sqrt{\frac{(\ell'-m \mp 1) !}{(\ell'+m \pm 1) !}}\,P_{\ell'}^{(m\pm 1)}\left(\cos(\theta_{v (t_r)})\right)\, e^{(m \pm 1)\phi''_{v(t_r)}},
\end{split}
\end{equation}
where $\mathbf{v}(t_r) = \mathbf{p}_s + \mathbf{A}(t_r)$ is the saddle point complex velocity vector at the time of recombination, whose radial, polar and azimuthal components are $v(t_r)$, $\theta_{v(t_r)}$ and $\phi_{v(t_r)}$, respectively. As pointed out above, the prime and
double prime superscripts indicate, respectively, the real and imaginary parts of the
corresponding quantity.

The radial factor $\mathcal{R}_{v(t_r),\ell'}$ is the same for $\hat{\mathbf{e}}_\pm$ and does not depend on the initial quantum number $m$, and hence does not influence the contrast between the lines (see Appendix \ref{app:IonRec}). The sum of index $\ell'$ runs over all angular momenta of the continuum states from which one-photon recombination occurs. Finally, the 3j-coefficients reflect standard angular momentum algebra for one-photon transitions. For recombination to an $s$ state, the 3j coefficients are zero unless $\ell'=1$, while for recombination to a $p$ state, they are zero unless $\ell'=0$ or $\ell'=2$.
The z-component of the recombination velocity is negligible for collinear bicircular fields,
so that $\cos \theta_{v(t_r)} = 0$. 

The difference between the recombination matrix element in Eq.~\eqref{eq:recombination_matrix_element} and  its inverse photoionization counter-part is
now clear. The inverse photo-absorption process occurs from a continuum
state with real velocity. The recombination process, due to the fact that the electron
is conditioned to return to the same place and state from which it was ionized, occurs
from a continuum state with complex velocity vector -- only its square has to be real-valued.

This difference leads to the appearance of the last exponential factor in Eq.~\eqref{eq:recombination_matrix_element}, which is absent in the inverse photo-absorption process.
We may thus identify two separate contributions to the recombination dipole.
The first is common with the photoionization process, and includes all of the terms
in Eq.~\eqref{eq:recombination_matrix_element} except for the last exponential. The second is
due to the recombination condition, which is given by the last exponential.
Each of these two contributions is governed by a different propensity rule.

\begin{figure}
\centering
\includegraphics[width=\linewidth]{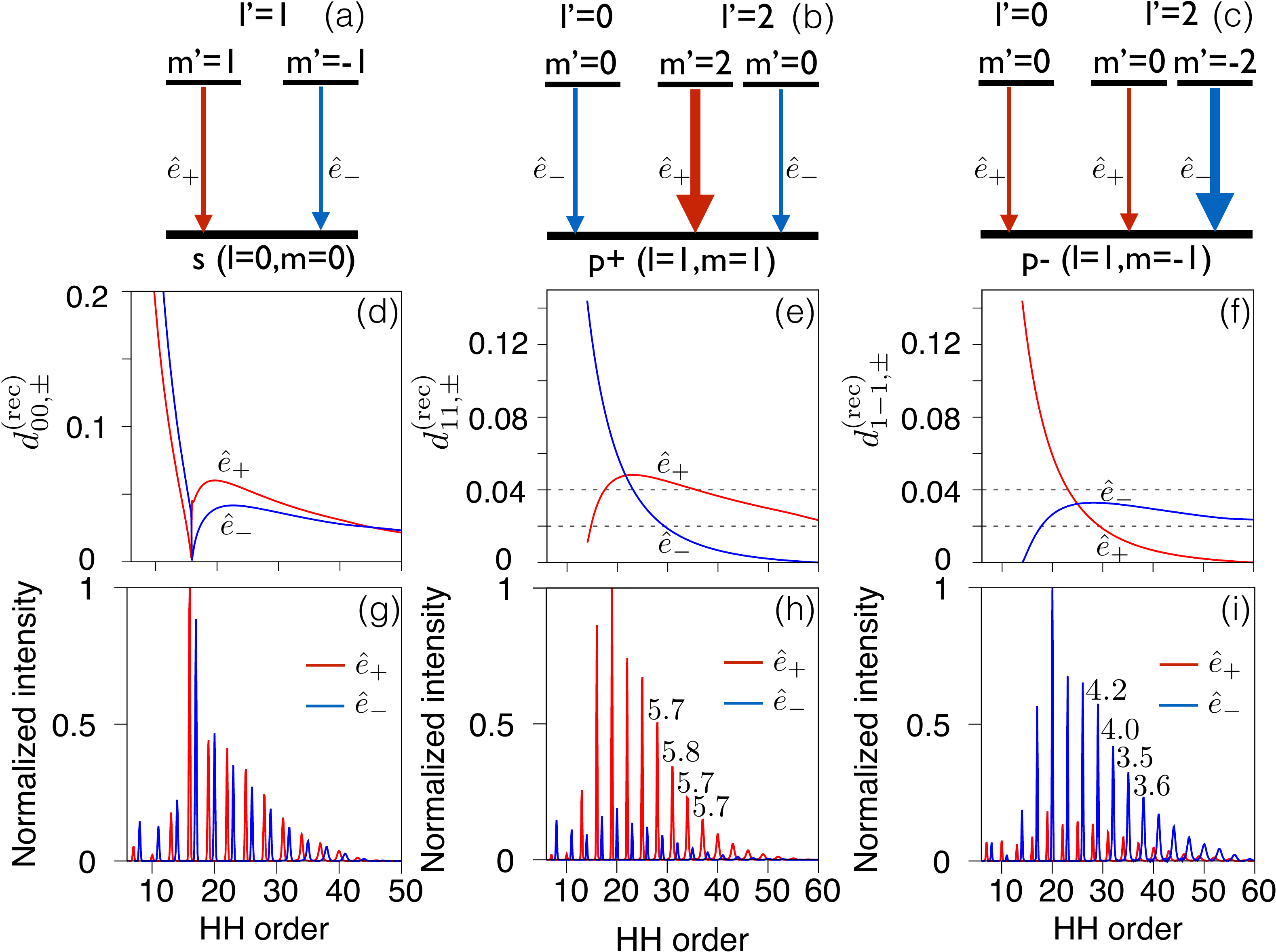}\caption{Top: permitted dipole transitions in recombination to (a) $s$, (b) $p+$ and (c) $p-$ states. Thick arrows indicate those favored by the Fano-Bethe propensity rule. Center: recombination matrix element (Eq.~\eqref{eq:recombination_matrix_element}), in arbitrary units, for final (d) $s$, (e) $p+$ and (f) $p-$ states. Bottom: Numerical TDSE high harmonic spectra when an electron is emitted from (g) $s$, (h) $p+$ and (i) $p-$ states. Colored lines indicate the helicity of the emitted light: red for $\hat{\mathbf{e}}_+$ (counter-clockwise) and blue for $\hat{\mathbf{e}}_+$ (clockwise).}\label{fig:Recombination}
\end{figure}

\subsection{Propensity 1: The Fano-Bethe rule.}

The first contribution, i.e.,  the emission analogue of the photo-absorption process, follows the propensity rule given by Fano and Bethe \cite{Fano1985, Bethe1964}. This states that in one-photon transitions, the helicity of the absorbed/emitted photon preferentially co-rotates with the final state (see Fig.~\ref{fig:Recombination}(a), Fig.~\ref{fig:Recombination}(b)). Mathematically, in part this is a consequence of
the fact that the 3-j coefficient in Eq.~\eqref{eq:recombination_matrix_element} is larger
when $m=1$ for $\hat{\mathbf{e}}_+$ and when $m=-1$ for $\hat{\mathbf{e}}_-$. In Figs.~\ref{fig:Recombination}(d) - \ref{fig:Recombination}(f))
we show the full recombination matrix element for
the $s$ ($\ell=0,m=0$), $p+$ ($\ell=1, m=1$) and $p-$ ($\ell=1,m=-1$) states, as
a function of the harmonic order.
To extract the recombination velocities, we solved the corresponding saddle point equations
and used the results for the wavefunction in Eq.~\eqref{eq:short_range_potential}.

For helium (Fig.~\ref{fig:Recombination}(d), Fig.~\ref{fig:Recombination}(g)), we considered an $s$-symmetric ground state with $I_p=0.9$~a.u. and driving fields of $12$~fs duration and field strengths $F_{\omega} = F_{2\omega} = 0.056$~a.u. For neon (Fig.~\ref{fig:Recombination}(e), Fig.~\ref{fig:Recombination}(f),Fig.~\ref{fig:Recombination}(h), Fig.~\ref{fig:Recombination}(i)), we considered a $p$-symmetric ground state with $I_p=0.797$~a.u. and driving pulses of $12$~fs duration and field strengths of $F_{\omega} = F_{2\omega} = 0.074$~a.u. The pulse parameters and $I_p$ were chosen to match those used in the TDSE calculations. 
As stated earlier, the $\hat{\mathbf{e}}_\pm$ component of the high harmonic lines are depicted as red/blue lines to highlight that they have the same helicity as that of the fundamental/second-harmonic driver.

For $p+$ (Fig.~\ref{fig:Recombination}(b), Fig.~\ref{fig:Recombination}(e),Fig.~\ref{fig:Recombination}(h)), the red lines co-rotate with the final state, while the blue lines counter-rotate with it. As predicted by the Fano-Bethe propensity rule, the red lines dominate the spectrum, with the exception of lower energies. The reason for the latter is simple. Recombination to a final state $p+$ happens from the two partial waves $s$ and $d$, as shown in Fig.~\ref{fig:Recombination}(a). While recombination from a $d$ partial wave can occur to states co-rotating and counter-rotating with the field,  recombination from an
$s$ wave is only possible to states counter-rotating with the field.
At lower energies, the probability of the electron recombining from an s-orbital is higher,
and this is reflected in the spectrum as a dominance of the counter-rotating (blue) lines.
For a final $p-$ state (Fig.~\ref{fig:Recombination}(c), Fig.~\ref{fig:Recombination}(f), Fig.~\ref{fig:Recombination}(i)), the argument is analogous to that for $p+$, only now the red lines counter-rotate with the final state, while the blue lines co-rotate with it.

It is worth noticing that the Fano-Bethe propensity rule cannot be modified by
altering the parameters of the laser pulses, since it only depends on the modulus of
the velocity, which is always fixed by the saddle point solutions,
i.e., $v(t_r) = \sqrt{2(N\omega - I_p)}$ \cite{Vrakking2014}.

\subsection{Propensity 2: The Recombination condition.}

If the recombination process in HHG were only the inverse of photoionization, then
the recombination matrix elements for states $p+$ and $p-$ would have been equivalent, up to
the corresponding change in the helicity of the emitted light.
A comparison of panels e and f in Fig.~\ref{fig:Recombination} shows that this is not the case.
While the counter-rotating lines are the same (blue for $p+$ and red for $p-$), the
co-rotating lines (red for $p+$ and blue for $p-$) are clearly different. The co-rotating line is more dominant in $p+$ than it is in $p-$.

This difference is the consequence of the recombination condition,  i.e.,
the factor $e^{(m\pm 1) \phi''_{v(t_r)}}$ in Eq.~\eqref{eq:recombination_matrix_element}.
This factor is equal to unity, and thus does not play a role
for the counter-rotating lines, but it enhances or suppresses
the co-rotating lines, depending on $\phi''_{v(t_r)}$.

If $\phi''_{v(t_r)}$ is positive, which is the case for the vast majority of relevant harmonic
orders and pulse parameters, then the red lines are enhanced in $p+$ while
the blue lines are damped in $p-$. Intuitively, this propensity rule accounts for
which photon is more likely to be absorbed or emitted in the free-free transition.
As mentioned in \cite{Dorney2017}, to lowest order, absorption of one extra photon of frequency $\omega$ (``red") leads to the
emission of the ``red" harmonic line. Absorption of one extra photon of frequency
2$\omega$ (``blue") leads to the emission of the ``blue" harmonic line.
For equal intensities of the $\omega$ and $2\omega$ fields, the low energy photons are more
likely to cause free-free transitions~\cite{Jimenez-Galan2016}, and thus the red lines dominate the harmonic spectrum.
In contrast to the Fano-Bethe propensity rule described above, this propensity rule (in particular $\phi''_{v(t_r)}$) depends on the parameters of the driving fields through the saddle
point equations, and thus offers excellent opportunities for control.

When the electron is emitted from a single orbital, e.g., $s$, $p+$ or $p-$, the sum in Eq.~\eqref{eq:dipole_reduced} runs over one single index, and the ionization term $\Upsilon_{\ell m} [\mathbf{p}_s + \mathbf{A} (t_i)]$ acts as a global factor. In this case, therefore,
only the two mentioned above propensity rules will be responsible for the contrast
between the harmonic lines in the spectrum. Indeed, in Figs.~\ref{fig:Recombination}(g) - Fig.~\ref{fig:Recombination}(i) we
show the solution of the TDSE for the case of ionization from $s$, $p+$ and $p-$ orbitals,
respectively. The contrast between the two helicities (red and blue lines) follows nicely
that predicted by our model, and thus follows the two simple propensity
rules highlighted above.
In particular, the numbers above the harmonic lines in Fig.~\ref{fig:Recombination}(h), Fig.~\ref{fig:Recombination}(i) depict the ratio between the maxima of consecutive 3N+1 and 3N+2 lines in $p+$
and between consecutive $3N+2$ and $3N+1$ lines in $p-$. The fact that red lines are stronger in $p+$ than
blue lines are in $p-$, is a manifestation of the recombination-conditioned propensity rule.

However, for all noble gas atoms except helium, the electron is emitted
from both the $p+$ and $p-$ orbitals. In this case, the ionization factor
now plays a crucial role. In essence, it acts as an additional weighting factor for the contribution
of $p+$ and $p-$ electrons. Indeed, in this case the $\hat{\mathbf{e}}_\pm$ components of the intensity
observed in the $p$ spectrum are proportional to
\begin{equation}\label{eq:intensity_2p}
\begin{split}
I_\pm &\propto \left| \langle \psi_{11} | \text{d}_\pm | \mathbf{v} (t_r) \rangle \right|^2  \left| \Upsilon_{11} [\mathbf{v} (t_i)] \right|^2 + \\
&+ \left| \langle \psi_{1-1} | \text{d}_\pm | \mathbf{v} (t_r) \rangle \right|^2 \left| \Upsilon_{1-1} [\mathbf{v} (t_i)] \right|^2  + \\
&+F_{int} \cos\left[ 2(\phi_{v(t_i)}' - \phi_{v (t_r)}') \right] e^{\pm 2\phi_{v (t_r)}''},
\end{split}
\end{equation}
where the first two terms above are the spectra observed when the
electron is emitted from the $p+$ and $p-$ orbitals, respectively, and
the last term is the interference term which is different for each helicity component $\hat{\mathbf{e}}_\pm$.
The phase of the interference term is composed of twice the phase difference
between the ionization and recombination angles, and thus depends on the pulse parameters
through the saddle point equations. However, the influence of the interference term on the contrast between the lines, while slightly enhancing the ``red" lines at low energies, is small
compared to the preceding two contributions in Eq.~\eqref{eq:intensity_2p} (see Fig.~\ref{fig:Ionization}). It is therefore not
a crucial term for our purposes and we will not comment on it further here.
We note, nonetheless, that if one wishes to go beyond the plane wave approximation and
consider the real scattering states, then the interference term will contain
the interplay between the scattering phases, which could lead to additional effects,
as pointed out in \cite{Baykusheva2016}.

Let us now concentrate on the first two terms in Eq.~\eqref{eq:intensity_2p} and pose the
question: which orbital has the strongest ionization factor, $p+$ or $p-$?
The answer
is given by the third propensity rule, described below.
Note that while the ionization factor depends on the complex-valued ionization
time $t_i$ and the velocity ${\bf v}(t_i)$, it is also conditioned on the electron
return to the parent core.

\subsection{Propensity 3: The Barth-Smirnova rule}

\begin{figure}
\centering
\includegraphics[width=\linewidth]{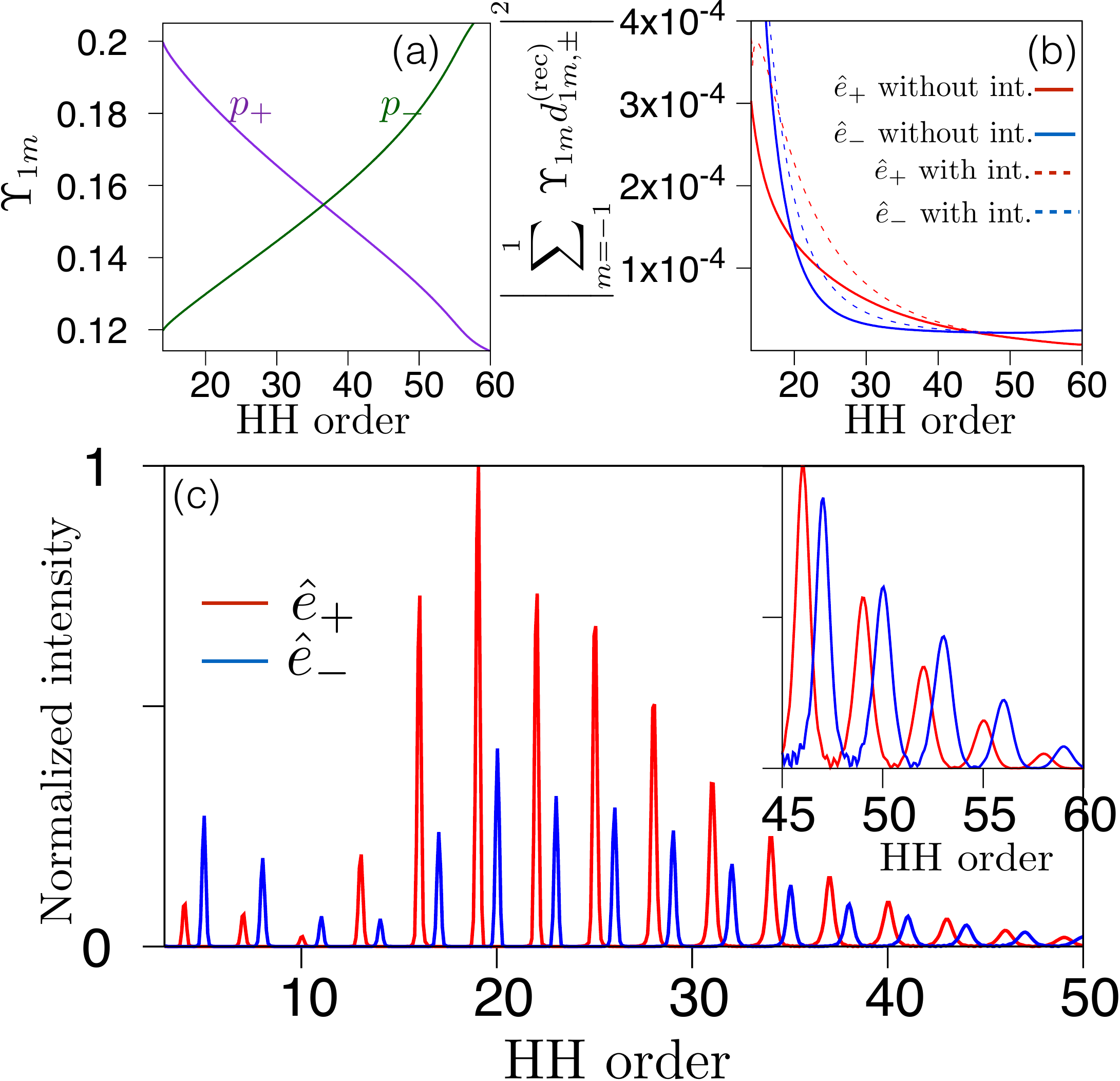}\caption{(a) SFA-calculated ionization factor, in arbitrary units, of electrons emitted from the $p+$ (magenta) and $p-$ (green) orbital of neon. (b) SFA-calculated square of the product of the ionization factor and recombination matrix element in the spectrum of neon separated in the two helical components (in arbitrary units), including (dashed lines) and neglecting (solid lines) the interference term in Eq.~\eqref{eq:intensity_2p}. (c) Spectrum of neon calculated solving the TDSE and separated in the two helical components (same as Fig.~\ref{fig:Spectra}(b)).}\label{fig:Ionization}
\end{figure}

Barth and Smirnova predicted that tunnel ionization
driven by circular fields preferentially removes electrons from states
counter-rotating with the field~\cite{Barth2011, Barth2013}.
The same rule applies in the bicircular case \cite{ayuso2017attosecond, Milosevic2016}, when
the electron is required to return to the core.
Incidentally, we note that this tunneling propensity rule is opposite to the
photoionization one given by Fano-Bethe's propensity rule. In Fig.~\ref{fig:Ionization}(a)
we show the ionization factor, as a function of the harmonic number,
for the electrons emitted from $p+$ and $p-$ states, and for the same pulse parameters
as those used in Fig.~\ref{fig:Recombination}.
The $p+$ state, which counter-rotates with respect to the total bi-circular field,
is dominant over the largest part of the relevant (more intense)
part of the spectrum. The ionization factor, however, depends on the energy,
and close to harmonic 40, emission from $p-$ state becomes dominant.
Again, we stress that this propensity rule depends on the pulse parameters, and
can therefore be altered.

To calculate the contrast of red to blue lines in the $p$ spectra within our model, we multiply the
$\hat{\mathbf{e}}_+$ component of the recombination factor in the $p+$ spectrum (red line in Fig.~\ref{fig:Recombination}(e)) by the $p+$ ionization factor (magenta line in Fig.~\ref{fig:Ionization}(a)), and sum it with the $\hat{\mathbf{e}}_+$ component of the recombination factor in the $p-$ spectrum (red line in Fig.~\ref{fig:Recombination}(f)) times the $p-$ ionization factor (green line in Fig.~\ref{fig:Ionization}(a)), according to Eq.~\eqref{eq:intensity_2p}. This gives us the amplitude of the $\hat{\mathbf{e}}_+$ (red) component in the $p$ spectrum. Analogously, we obtain the $\hat{\mathbf{e}}_-$ (blue) component.
The $\hat{\mathbf{e}}_+$ and $\hat{\mathbf{e}}_-$ amplitudes obtained this way are plotted in Fig.~\ref{fig:Ionization}(b),
as a function of the harmonic order. As commented earlier, Fig.~\ref{fig:Ionization}(b) shows that neglecting the interference term
(last term in Eq.~\eqref{eq:intensity_2p}) gives essentially the same ``red" and ``blue" contrast, except at lower orders, where SFA is not accurate in the first place. The red and blue lines in the spectrum calculated from the TDSE (Fig.~\ref{fig:Ionization}(c))
follows that predicted by the model (square of the solid red and blue lines in Fig.~\ref{fig:Ionization}(b)) and coincides also with that predicted by the SFA in the rotating frame of reference~\cite{Pisanty2017}.

We now have all the ingredients to understand the features in the HHG spectrum of
neon, which we reproduce again in Fig.~\ref{fig:Ionization}(c) for clarity. The
spectrum shows a clear dominance of the red lines in the plateau region, as our
model predicts. For increasing harmonic orders, the blue lines become comparable to
the red lines and, near the cut-off region ($\simeq$ HH45), the blue lines dominate.
This feature is also well reproduced by our simple model. For low-energy harmonics
the agreement is not so good, which is expected since the strong field approximation is
not suited to treat below-threshold and close-to-threshold harmonics.
Nonetheless, the higher contribution of blue lines at harmonics close to threshold,
due to the stronger contribution of the $s$-wave to the
recombination process as highlighted earlier, is observed in both the spectrum and our model.

\section{Control of the helicity and harmonic contrast in bi-circular fields}

In the previous section we have outlined the three propensity rules
responsible for the relevant features in the spectra. In this section we will use
them to exert control over the ratio between the red and blue harmonic lines.
As we have pointed out, the Fano-Bethe propensity rule cannot be altered by changing
the pulse parameters, but we will now show how the recombination and
ionization propensity rules modify the spectrum when the intensities of the two pulses are varied.

\begin{figure}
\centering
\includegraphics[width=\linewidth]{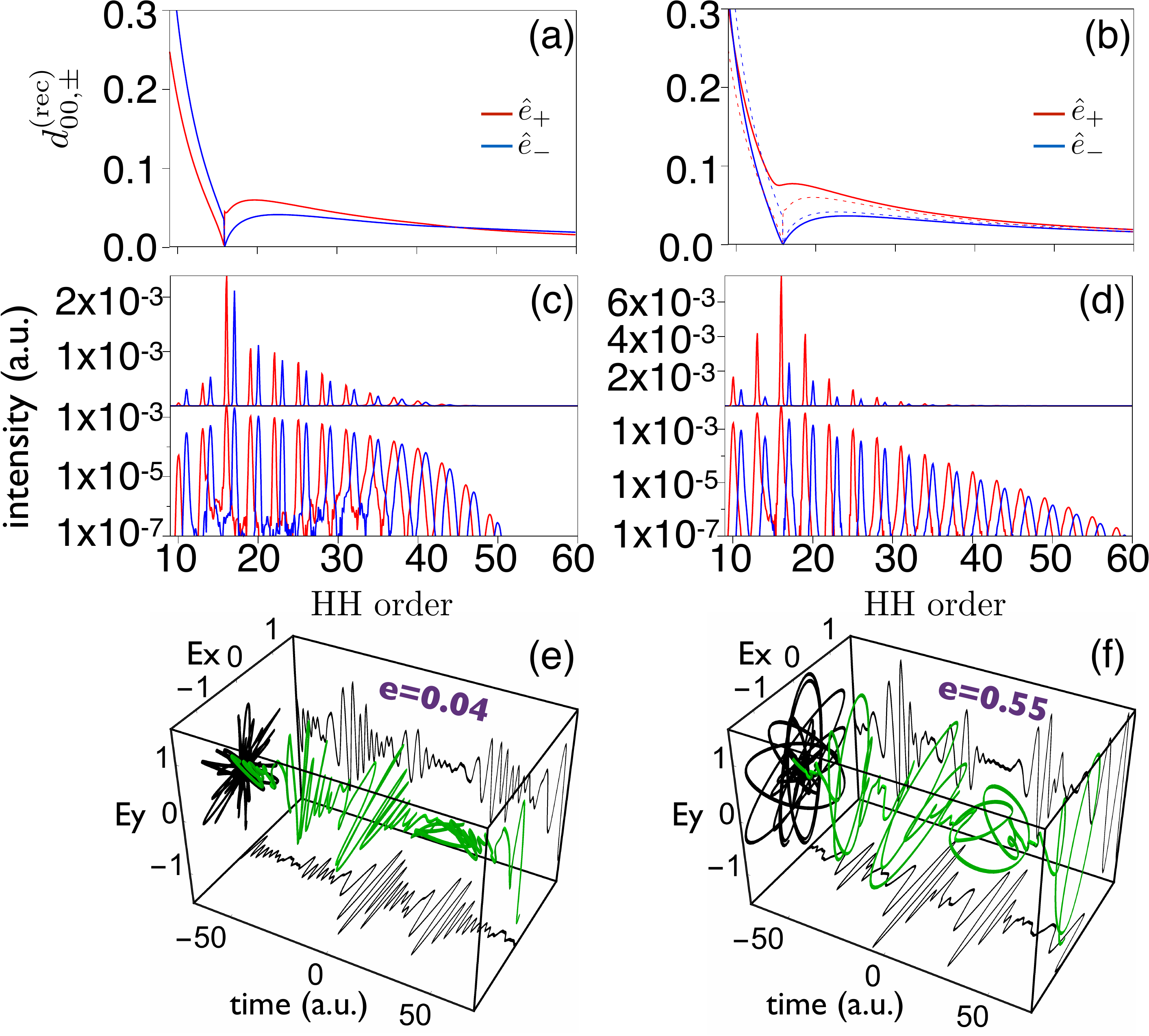}\caption{Control of helicity of
attosecond pulses generated in Helium, using the bicircular driving field.
Each of the two columns represents a different intensity ratio between the fundamental and
the second harmonic: $I_{\omega}/I_{2\omega}=1$ for panels (a,c), and
$I_{\omega}/I_{2\omega}=2.25$ for panels (b,d).
Top row: SFA-calculated recombination matrix element for helium, as a function of harmonic order, in arbitrary units. Center row: TDSE-calculated spectrum of helium in linear (top) and logarithmic (bottom) scale. Bottom row:
attosecond pulse trains (APT) generated by taking an energy window from harmonic 9 to harmonic 42. The degree of circularity $e$ of the generated APTs is given (see text for details) and electric field strengths are expressed in arbitrary units. The pulse parameters used in both the SFA and TDSE calculations are the same.}\label{fig:SpectrumHelium}
\end{figure}

Let us first consider helium. As stated previously, in this case
the ionization factor does not play any role in the contrast between the two helicities, so
this contrast gives us direct access to the propensity rules in the recombination process.
Fig.~\ref{fig:SpectrumHelium}(a), Fig.~\ref{fig:SpectrumHelium}(b) shows the recombination factor
of helium,
$\langle \psi_{0 0} | \hat{\mathbf{d}} \cdot \hat{\mathbf{e}}^*_\pm | \mathbf{v} (t_r) \rangle$,
for two intensity ratios between the fundamental and second harmonic,
$I_{\omega}/I_{2\omega}=1$ and $2.25$, respectively. The
duration of both driving pulses was set to $12$~fs and the field strength of the second harmonic
was always fixed at $F_{2\omega}=0.056$~a.u. As the ratio increases,
so does the contrast between the red and blue lines in both the recombination amplitude
and in the HHG spectrum (Fig.~\ref{fig:SpectrumHelium}(b), Fig.~\ref{fig:SpectrumHelium}(d)). Higher red intensities translate
into a higher number of ``red" photons to be absorbed or emitted in the
continuum and, consequently, to a stronger dominance of red harmonic lines in the
spectrum.

In Fig.~\ref{fig:SpectrumHelium}(e), Fig.~\ref{fig:SpectrumHelium}(f) we show the attosecond pulse train (APT)
computed by inverse Fourier transform of the corresponding spectra, taking an energy window from harmonic 9 up to harmonic 42.
When the intensity of the second harmonic driver is the same as that of the fundamental, there are three linearly polarized bursts per laser cycle ($T \approx 110$~a.u).
When the intensity of the fundamental is 2.25 that
of the second harmonic (Fig.~\ref{fig:SpectrumHelium}(f)), the bursts from the APT are highly
elliptically polarized and rotating in the $\hat{\mathbf{e}}_+$ (counter-clockwise) direction. We can calculate the degree of circularity of the generated attosecond pulse
train by integrating the two helical components $E_{\pm} = \mp (E_x\pm iE_y)/\sqrt{2}$ of the electric field over a temporal window, which we choose from $t_1=-70$~a.u. to $t_2 = 70$~a.u. 
(same interval as that shown in Fig.~\ref{fig:SpectrumHelium}(e), Fig.~\ref{fig:SpectrumHelium}(f) ).
The degree of circularity can then be defined as $e=(|E_+|^2 - |E_-|^2)/(|E_+|^2 + |E_-|^2)$.
Positive (negative) $e$ will yield an attosecond pulse train elliptically polarized in the counter-clockwise (clockwise) direction; $e=0$ corresponds to a linearly polarized field, while $|e|=1$ corresponds to a circularly
polarized field. The value of $e$ for the corresponding generated APT is shown in Fig.~\ref{fig:SpectrumHelium}(e), Fig.~\ref{fig:SpectrumHelium}(f). We can control
the polarization by simply changing the intensity of the red driving field with
respect to the second harmonic. Note that in this case, the generated bursts from the APT will always have the same helicity as the fundamental driving field,
regardless of the frequency window applied to the spectrum to obtain them.

It is important to stress that the contrast between red and blue lines in helium appears because of the recombination-conditioned propensity rule, i.e., due to the fact that the trajectories
propagate in complex space-time. If the electron trajectories moved in real-valued space-time,
then no contrast would have been observed, and the APT generated in helium will
always be linearly polarized, irrespective of the intensity ratio between the
fundamental driver and the second harmonic.

\begin{figure}
\centering
\includegraphics[width=\linewidth]{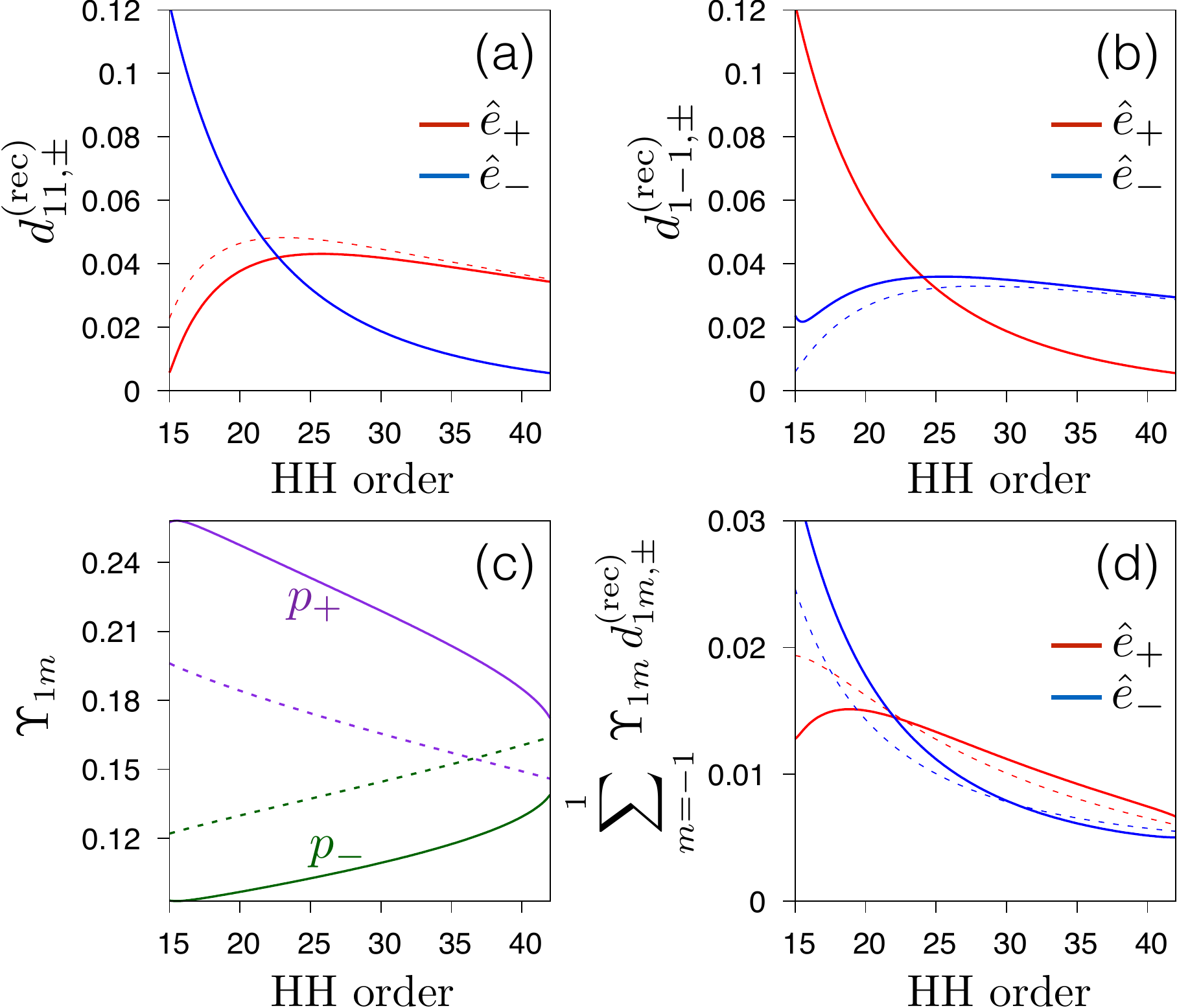}\caption{
Top row: SFA-calculated matrix elements for recombination to the
(a) $p+$ (b) $p-$ orbitals of neon in arbitrary units.
Bottom row: (c) SFA-calculated ionization factor of electrons emitted from the
$p+$ (magenta) and $p-$ (green) orbitals of neon and (d) SFA-calculated product of
the ionization and recombination dipoles in the spectrum of neon, in arbitrary units. For all panels, solid red (blue) lines indicate emission
of light with $\hat{\mathbf{e}}_+$ ($\hat{\mathbf{e}}_-$) helicity using a driver intensity
ratio of $I_{\omega}/I_{2\omega} = 0.5$. The dashed red (blue)
lines indicate emission of light with $\hat{\mathbf{e}}_+$ ($\hat{\mathbf{e}}_-$) helicity using a driver
intensity ratio of $I_{\omega}/I_{2\omega}= 1$.}\label{fig:RecIon_2wstronger}
\end{figure}

\begin{figure*}
\centering
\includegraphics[width=\linewidth]{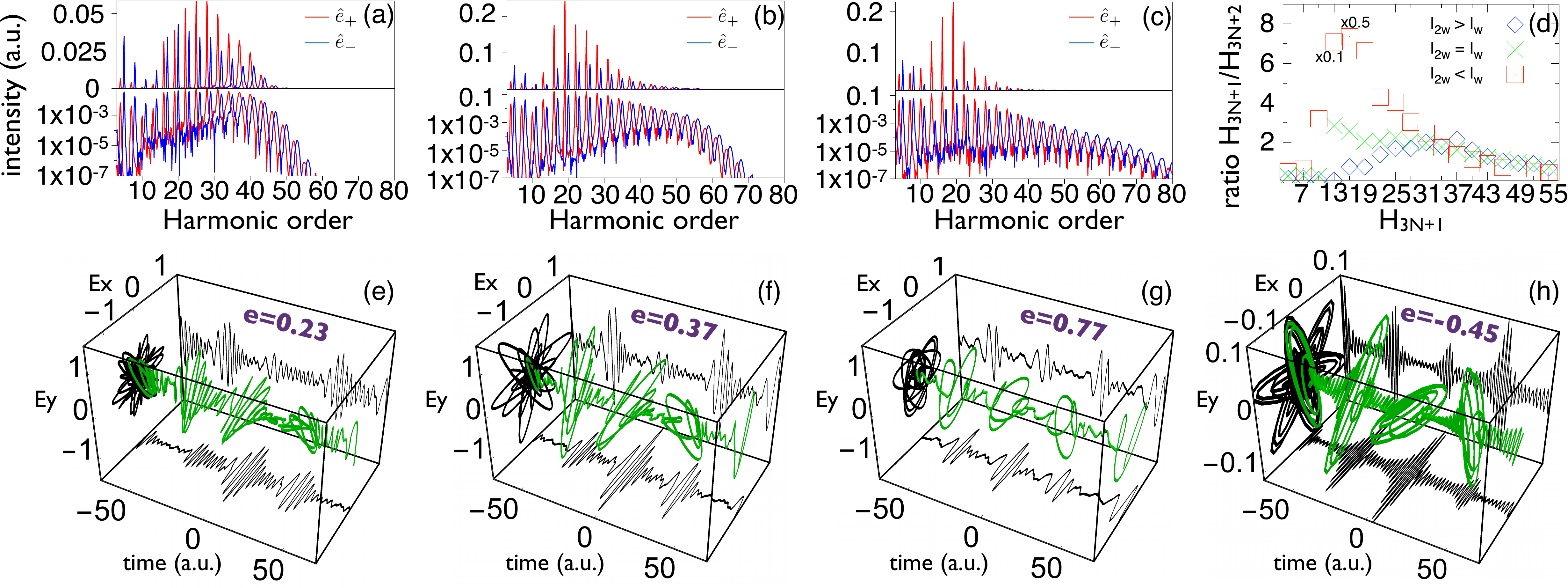}\caption{Control of helicity of
attosecond pulses generated in Neon, using the bicircular driving field. TDSE-calculated HHG spectra in neon (left panels, with linear scale in the top and logarithmic scale in the bottom) and APT obtained from the spectra (right panels) for three different driver intensity ratios: (a,e) $I_{\omega}/I_{2\omega} = 0.5$, (b,f) $I_{\omega}/I_{2\omega} = 1$, (c,g,h) $I_{\omega}/I_{2\omega} = 2$. The APTs in panels (e,f,g) were obtained by inverse Fourier transform of the plateau harmonics in panels (a,b,c), respectively, while the APT in panel (h) was obtained from the cut-off harmonics of panel (c). Electric field strengths are expressed in arbitrary units. Panel (d) shows the ratio between the peak intensity of $H_{3N+1}$ and $H_{3N+2}$ for an $I_{\omega}/I_{2\omega}$ ratio of 0.5 (blue diamonds), 1 (green crosses) and 2 (red squares). A factor of 0.1 and 0.5 has been applied to $H_{13}/H_{14}$ and $H_{16}/H_{17}$, respectively, in the case of $I_{\omega}/I_{2\omega} = 2$. The black line indicates equal intensity of $H_{3N+1}$ and $H_{3N+2}$ harmonics.}\label{fig:SpectrumNeon}
\end{figure*}

We now turn our attention to neon. In Fig.~\ref{fig:RecIon_2wstronger}
we show the recombination factor for $p+$ and $p-$ (panels a and b),
the ionization factor (panel c), and the resulting ratio between red and blue harmonic lines in the spectrum (panel d), when $I_{\omega}/I_{2\omega} = 0.5$.
In this case, there are more ``blue" photons to be exchanged with the field during the free-free transitions,
and hence the blue lines would be stronger than in the case of equal intensities.
This is what we observe in the recombination factor,
shown in Fig.~\ref{fig:RecIon_2wstronger}(a), Fig.~\ref{fig:RecIon_2wstronger}(b) for the $p+$ and $p-$ spectra, respectively. The ionization factor now favors the $p+$ spectrum throughout the below-threshold and plateau
harmonics (Fig.~\ref{fig:RecIon_2wstronger}(c) ). The $p+$ spectrum, however, has
extended the dominance of the blue lines to higher
harmonic orders. This, added to a decrease in the red/blue ratio, is detrimental to generating highly circular APT in the plateau.
Additionally, increasing the intensity of the second harmonic increases the contribution of forbidden harmonics due to Rydberg state population, as seen in Figs.~\ref{fig:NeSp-1},\ref{fig:NeSp-2} and also in~\cite{Jimenez-Galan2017}.
In this case, the small radial box used to reduce computational time and the influence of long trajectories, eliminates the contribution of Rydberg orbits.
The predicted contrast is given in Fig.~\ref{fig:RecIon_2wstronger}(d), which is in
good agreement with what the TDSE spectrum shows (Fig.~\ref{fig:SpectrumNeon}(a) ).
The pulses in this case were 12~fs long with a field strength of $F_{\omega} = F_{2\omega}/\sqrt{2} = 0.052$~a.u.
The bursts of the APT generated in this case by filtering everything but the plateau region are close to linear ($e=0.23$, Fig.~\ref{fig:SpectrumNeon}(e) ).

\begin{figure}
\centering
\includegraphics[width=\linewidth]{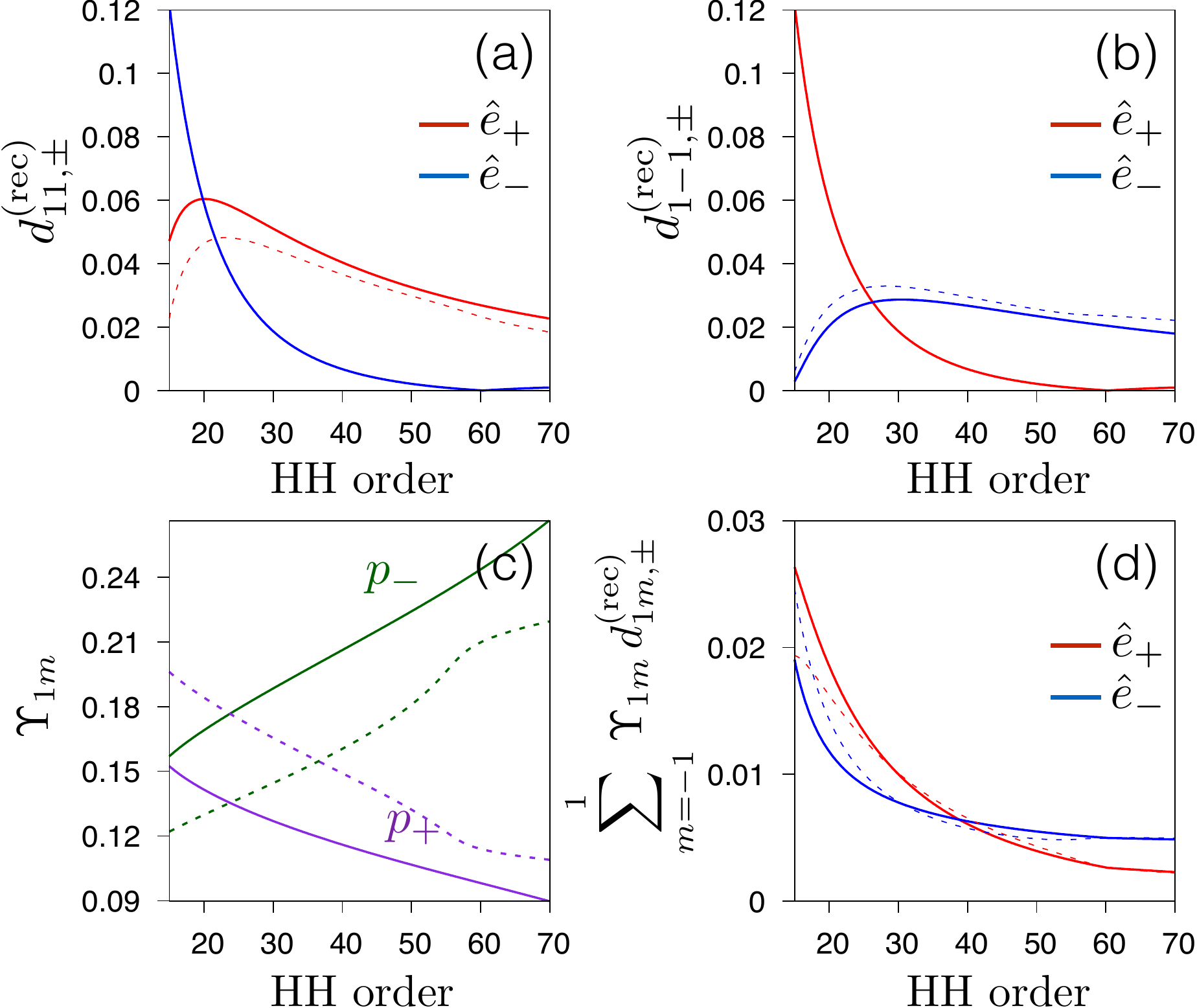}\caption{Same as Fig.~\ref{fig:RecIon_2wstronger}, but solid lines now indicate results using a driver intensity ratio of $I_{\omega}/I_{2\omega} = 2$.}\label{fig:RecIon_wstronger}
\end{figure}

What happens when the fundamental field is stronger than the second harmonic?
Again, we see a
stronger dominance of red harmonic lines in the recombination factor in
both the $p+$ and $p-$ spectra due to the higher number of ``red" photons
exchanged in the continuum, as we saw in helium.
As for the ionization factor, the $p-$ spectrum now dominates over all the relevant harmonic region.
The latter is responsible for an extended dominance of red lines to lower harmonic orders, as it is illustrated in Fig.~\ref{fig:RecIon_wstronger}(d). The interplay between the ionization and recombination factors leads to a higher and more extended red/blue contrast in the spectrum in this case, as compared to the case of equal intensities. We observe this in both our model prediction (Fig.~\ref{fig:RecIon_wstronger}(d) ) and the HHG spectrum (Fig.~\ref{fig:SpectrumNeon}(c) ), both computed with the same pulse parameters. In particular, we used a field strength of $F_{\omega} = \sqrt{2} F_{2\omega} = 0.1$~a.u.
Therefore, just like in the case of helium, increasing the fundamental intensity
with respect to the second harmonic dramatically increases the circularity
of the APT generated with the plateau harmonics to the value $e=0.77$, which we show in Fig.~\ref{fig:SpectrumNeon}(g).
In Fig.~\ref{fig:SpectrumNeon} (d) we show the intensity ratio of the consecutive harmonics $H_{3N+1}/H_{3N+2}$ for three different $I_{\omega}/I_{2\omega}$ ratios: 0.5 (blue diamonds), 1 (green crosses) and 2 (red squares). The ratio between consecutive harmonics clearly increases as the $I_{\omega} / I_{2\omega}$ ratio is increased for the most intense high harmonics.
Furthermore, the higher total intensity in this case allows us to look at the cut-off harmonics more clearly. The change from dominating ``red" lines to ``blue" lines in this region is evident, as also predicted by our model (see Fig.~\ref{fig:RecIon_wstronger}(d) ). In Fig.~\ref{fig:SpectrumNeon}(h) we show the APT obtained from filtering out all harmonics below HH48. The bursts are remarkably elliptical and very clearly separated. The helicity of the bursts in this case is opposite to those obtained by filtering the plateau energies only, i.e., the bursts rotate in the $\hat{\mathbf{e}}_-$ direction, with a degree of circularity of $e=-0.45$. Only by spectral filtering, therefore, we can, within a single experiment, obtain two circular attosecond pulse trains rotating in opposite directions.

\section{Conclusions} \label{section_conclusions}
In conclusion, we have developed comprehensive analytical understanding of the
electron  dynamics occurring in the high harmonic generation process driven
by two-color counter-rotating fields, and have linked these dynamics to the
features observed in the experimental spectrum. We have identified the three
propensity rules responsible for the contrast between the 3N+1 and 3N+2 harmonic
lines in the HHG spectra of noble gas atoms, and demonstrated how these
rules depend on the laser parameters and can be used to shape the
polarization properties of the emitted attosecond pulses.

In particular, we were able to show that for atoms emitting from an $s$-shell (e.g., helium),
the contrast between the two light helicities in the HHG spectra
follows the one-photon recombination matrix element but still encodes the
condition of the electron return. This offers one a
unique opportunity to obtain  the characteristics of
what is, formally, recombination from a continuum state with complex-valued
velocity.

For atoms emitting $p$ electrons, information on the different ionization rates of $p+$ and $p-$ electrons can similarly be obtained. This understanding of the dynamics demonstrates an easy to implement and efficient mechanism to control the polarization of the generated attosecond pulse trains. By increasing the ratio of intensities between the fundamental and second harmonic drivers, the contrast between the two adjacent harmonic lines in the spectrum dramatically increases, leading to more circular attosecond bursts. Moreover, we have showed that, for $p$ states, when the fundamental field is stronger than the second harmonic, the APT generated from the plateau harmonics rotates with the fundamental driver, while that generated by the cut-off harmonics rotates with the second harmonic. In this way, we are able to obtain, from the same experiment, two circularly polarized APTs with opposite helicity. Contrary to the common view, we have shown that it is possible to generate highly chiral attosecond bursts by using initial $s$-orbitals. Furthermore, we have linked this possibility to the fact that the electron carries complex velocity at the time of recombination, and thus represents a measurable confirmation of the electron's propagation in complex space-time during the HHG process.
\section{Acknowledgements}
AJG, NZ, and MI acknowledge financial support from the DFG QUTIF grant IV 152/6-1. 
DA and OS acknowledge support from the DFG grant SM 292/2-3.
EP acknowledges financial support from MINECO grants
FISICATEAMO (FIS2016-79508-P) and Severo Ochoa (SEV-2015-0522), Fundaci\'o Cellex, Generalitat de Catalunya (2014 SGR 874 and CERCA/Program), and ERC grants EQuaM (FP7-ICT-2013-C No. 323714), QUIC (H2020-FETPROACT-2014 No. 641122) and OSYRIS (ERC-2013-ADG No. 339106).

\appendix

\section{Relations between velocity components in the saddle point method}

It is useful to derive some relations between the $x$ and $y$ components of the velocities that come out of the ionization and recombination saddle points, i.e., 
\begin{equation}\label{eq:SaddlePointVelocity}
v(t_i)^2 = -2I_p, \quad \text{and} \quad v(t_r)^2=2(N\omega-I_p).
\end{equation}
In general, both ionization and recombination velocities are complex,
\begin{equation}
v^2 = v_x'^2-v_x''^2+v_y'^2-v_y''^2+2i (v_x'v_x'' + v_y'v_y'').
\end{equation}
Using the saddle point equations \ref{eq:SaddlePointVelocity}, both ionization and recombination velocities have thus the fixed relation
\begin{equation}\label{eq:RelationVelocities}
v_x'v_x'' = -v_y'v_y''.
\end{equation}
Using \eqref{eq:RelationVelocities}, one can check that, since the square of the ionization velocity needs to be negative according to its saddle point equation, necessarily
\begin{equation}\label{eq:RelationVelocities_ion}
v_y''(t_i)^2  > v_x'(t_i)^2.
\end{equation}
Analogously, for the recombination velocity we must have
\begin{equation}\label{eq:RelationVelocities_rec}
\begin{split}
&v_y''(t_r)^2 > v_x'(t_r)^2, \quad \text{for below threshold harmonics},\\
&v_y''(t_r)^2 < v_x'(t_r)^2, \quad \text{for above threshold harmonics}, \\
&v_y''(t_r)^2 = v_x'(t_r)^2, \quad \text{at threshold}.
\end{split}
\end{equation}
Finally, we point out that, due to the symmetry of our field, both the real and imaginary parts of the ionization and recombination velocities of consecutive bursts transform following a clockwise rotation of $2\pi/3$, i.e.,
\begin{equation}\label{eq:VelocityTransformations}
\begin{split}
&v_x^{(2)} = v_x^{(1)} \cos(2\pi/3) + v_y^{(1)}\sin(2\pi/3) \\
&v_y^{(2)} = -v_x^{(1)} \sin(2\pi/3) + v_y^{(1)}\cos(2\pi/3).
\end{split}
\end{equation}

\section{Expressions for real and imaginary velocity angles}
\label{app:velocityangles}
The saddle point analysis yields complex kinetic momentum $v_x$ and $v_y$, which can be expressed in complex spherical coordinates $(v, \theta, \phi)$. The module of this velocity is fixed by the saddle point equation as highlighted in Appendix I. From \eqref{eq:SaddlePointVelocity} one can see that the module of the velocity in ionization and in recombination for below threshold harmonics is imaginary. Hence, for ionization and recombination to below threshold harmonics, one can use the expansion of the trigonometric functions of complex arguments and write
\begin{equation}\label{eq:Cartesian2Spherical}
\begin{split}
&v_z = iv \cos\theta = 0, \\
&v_x' + iv_x'' = i v \cos\phi = i v \left[ \cos \phi' \cosh \phi'' - i\sin\phi' \sinh \phi''\right], \\
&v_y' + iv_y'' = iv \sin \phi = iv \left[\sin\phi' \cosh \phi'' + i\cos\phi' \sinh\phi''\right],
\end{split}
\end{equation}
where now $v$ is real, and is fixed up to a sign. We now identify the real and imaginary parts,
\begin{equation}\label{eq:vx'}
v_x'=v\sin\theta' \sinh\phi''
\end{equation}
\begin{equation}\label{eq:vx''}
v_x'' = v\cos\phi'\cosh\phi''
\end{equation}
\begin{equation}\label{eq:vy'}
v_y' = -v\cos\phi' \sinh\phi''
\end{equation}
\begin{equation}\label{eq:vy''}
v_y'' = v\sin\phi'\cosh\phi''.
\end{equation}
To obtain the real part of $\phi$ we can use, for example, \eqref{eq:vx''} and \eqref{eq:vy''},
\begin{equation}\label{eq:real1}
\phi' = \arctan \frac{v_y''}{v_x''},
\end{equation}
or \eqref{eq:vx'} and \eqref{eq:vy'},
\begin{equation}\label{eq:real2}
\phi'=\arctan \frac{-v_x'}{v_y'}.
\end{equation}
For the imaginary part we can use for example \eqref{eq:vx'} with \eqref{eq:vy''},
\begin{equation}\label{eq:im3}
\phi''=\arctanh \frac{v_x'}{v_y''},
\end{equation}
or \eqref{eq:vy'} with \eqref{eq:vx''},
\begin{equation}\label{eq:im4}
\phi''=\arctanh \frac{-v_y'}{v_x''}.
\end{equation}
Indeed, the different choices are a consequence of the relations \eqref{eq:RelationVelocities}. The apparent overdetermination is not such, since the choice we make here will automatically fix the sign of $v$, which is also an unknown. Indeed, if we use \eqref{eq:real1} for the real part and \eqref{eq:im3} for the imaginary part, then
\begin{equation}
v=-\frac{v_y'}{\cos\phi' \sinh\phi''},
\end{equation}
while if we use \eqref{eq:real1} and \eqref{eq:im4}, then
\begin{equation}
v=\frac{v_x'}{\sin\phi' \sin\phi''}.
\end{equation}
Equivalently, we can also choose \eqref{eq:real2} and \eqref{eq:im3}, so that
\begin{equation}
v=\frac{v_x''}{\cos\phi' \cosh\phi''},
\end{equation}
or \eqref{eq:real2} with \eqref{eq:im4}, and
\begin{equation}
v=\frac{v_y''}{\sin\phi' \cosh\phi''}.
\end{equation}
Of course, regardless of the choice, the result is the same.

For recombination to above threshold harmonics, the same argument follows, only now the imaginary number need not be present in \eqref{eq:Cartesian2Spherical} since the module of the velocity is already real. For example, one can choose to define
\begin{equation}
\begin{split}
&\phi' = \arctan \frac{v_y'}{v_x'}, \\
&\phi''= \arctanh \frac{v_y''}{v_x'}, \quad \text{and} \\
&v = \frac{-v_x''}{\sin\phi' \sinh\phi''}.
\end{split}
\end{equation}

\section{Ionization factor and recombination matrix element}
\label{app:IonRec}
The explicit steps of the derivation for the ionization factor can be found in~\cite{Barth2013}, while that for the recombination matrix element directly follows from the same procedure. Here we just reproduce the results.
For the ionization factor we have
\begin{equation}
\Upsilon [\mathbf{v} (t_i)] = \left[ \frac{(\mathbf{v}(t_i))^2}{2} + I_p \right] \langle \mathbf{v}(t_i) | \varphi_{\ell m} (\mathbf{r}) \rangle = \mathcal{F}_{ion}\,e^{im\phi_{v(t_i)}},
\end{equation}
where the factor
\begin{equation}
\mathcal{F}_{ion} = C_{\kappa \ell} \sqrt{\frac{2\kappa}{\pi}} \left(\frac{v(t_i)}{\kappa}\right)^{\ell}  \sqrt{\frac{2\ell + 1 (\ell - |m|)!}{4\pi (\ell + |m|)!}} P_{\ell}^{|m|} (\cos \theta_v),
\end{equation}
is independent on the sign of $m$. The complex angle $\phi_{v(t_i)}$ has the following real and imaginary components (see Appendix \ref{app:velocityangles})
\begin{equation}
\phi'_{v(t_i)} = \text{atan} \frac{v_y''(t_i)}{v_x''(t_i)}, \quad \phi''_{v(t_i)} = \text{atanh} \frac{v_x'(t_i)}{v_y''(t_i)}.
\end{equation}
With the definition of the inverse hyperbolic tangent, $\text{atanh}(z) = 1/2 \log \frac{1+z}{1-z}$, we can write the absolute value of the ionization step as
\begin{equation}\label{eq:IonizationAmplitude}
| \Upsilon [\mathbf{v} (t_i)]  | =| \mathcal{F}_{ion} | \left(\frac{v_y''(t_i) + v_x'(t_i)}{v_y''(t_i)-v_x'(t_i)}\right)^{-m/2},
\end{equation}
and the argument as
\begin{equation}
\arg \Upsilon [\mathbf{v} (t_i)]  = m\,\text{atan} \frac{v_y''(t_i)}{v_x''(t_i)} + \arg\mathcal{F}_{ion}.
\end{equation}
Using \eqref{eq:IonizationAmplitude} along with \eqref{eq:RelationVelocities} and \eqref{eq:VelocityTransformations}, one finds that this absolute value is equal for the three bursts, so
\begin{equation}\label{eq:SameAbsIon}
\frac{| \Upsilon [\mathbf{v} (t_i^{(2)}) ] |} { | \Upsilon [\mathbf{v} (t_i ^{(1)})] |}  = \frac{| \Upsilon [\mathbf{v} (t_i ^{(3)})] |} { | \Upsilon [\mathbf{v} (t_i ^{(1)})] |} = 1.
\end{equation}
We point out that $\mathcal{F}_{ion}$ and, in particular, $v(t_i)$, does not change from burst to burst since it is fixed by the saddle point equation ($v(t_i) = \sqrt{-2I_p}$). 
Similarly, using \eqref{eq:VelocityTransformations} along with the property $\text{atan} \left( \frac{x+y}{1-xy} \right) = \text{atan} (x) + \text{atan} (y)$, we obtain
\begin{equation}\label{eq:SameArgIon}
\begin{split}
&\arg \left(\frac{\Upsilon [\mathbf{v} (t_i ^{(2)})]}{\Upsilon [\mathbf{v} (t_i ^{(1)})]}\right) = 2m\pi/3, \\
&\arg \left(\frac{\Upsilon [\mathbf{v} (t_i^{(3)})]}{\Upsilon [\mathbf{v} (t_i ^{(1)})]}\right) = 4m\pi/3.
\end{split}
\end{equation}

Following analogous steps as those given in~\cite{Barth2013}, the recombination matrix element can be written as
\begin{equation}
\text{d}_{\ell,m, \pm} ^{\text {(rec)}} [\mathbf{v}(t_r)] = \mathcal{F}_{rec}\,e^{-i(m'\pm 1)\phi_{v(t_r)}},
\end{equation}
where, using Eq.~\eqref{eq:short_range_potential},
\begin{equation}\label{eq:recfactor}
\begin{split}
&\mathcal{F}_{rec} = (-1)^{m\mp 1}\,\frac{C^*_{\kappa \ell}}{\sqrt{8\pi\,\kappa^5}}\,\sqrt{(2\ell+1)} \times \\
& \times \sum_{\ell'=0}^{\infty}\,(2\ell'+1)\,\tj{\ell}{\ell'}{1}{0}{0}{0} \tj{\ell}{\ell'}{1}{-m}{(m\pm 1)}{\mp 1}  \times \\
&\times \sqrt{\frac{(\ell'-m \mp 1) !}{(\ell'+m\pm 1) !}}\,P_{\ell'}^{(m\pm 1)}\left(\cos(\theta_v (t_r))\right) \Gamma(\ell'+3) \times \\
&\times \left(\frac{i v}{2\kappa}\right)^{\ell'}\,_2F_1\left(\ell'/2+3/2, \ell'/2+2; \ell'+3/2; -\frac{v^2}{\kappa^2}\right).
\end{split}
\end{equation}
The real and imaginary components of the complex angle $\phi_{v (t_r)}$ are
\begin{equation}
\phi'_{v (t_r)} = \text{atan} \frac{v_y''(t_r)}{v_x''(t_r)}, \quad \phi''_{v(t_r)} = \text{atanh} \frac{v_x'(t_r)}{v_y''(t_r)},
\end{equation}
for below threshold harmonics, and
\begin{equation}
\phi'_{v(t_r)} = \text{atan} \frac{-v_x''(t_r)}{v_y''(t_r)}, \quad \phi''_{v(t_r)} = \text{atanh} \frac{v_y''(t_r)}{v_x'(t_r)}
\end{equation}
for above threshold harmonics. Again, with the definition of the inverse hyperbolic tangent we may write the absolute value of the recombination matrix element as
\begin{equation}
\begin{split}
&|\text{d}_{\ell, m, \pm} ^{\text {(rec)}} [\mathbf{v}(t_r)] | = | \mathcal{F}_{rec} | e^{(m'\pm 1) \phi''_{v(t_r)}} = \\
&= | \mathcal{F}_{rec} | (-1)^{\gamma(m'\pm 1)/2}\,\left(\frac{v_y''(t_r) + v_x'(t_r)}{v_y''(t_r)-v_x'(t_r)}\right)^{(m'\pm 1)/2},
\end{split}
\end{equation}
where $\gamma=0$ or $1$ depending on if we are considering below or above threshold harmonics. The expression for the argument of the recombination is
\begin{equation}
\arg \{ \text{d}_{\ell, m, \pm}^{\text {(rec)}} [\mathbf{v}(t_r)]  \}  = (m'\pm 1) \phi'_{v(t_r)} + \arg(\mathcal{F}_{rec}).
\end{equation}
Using the same arguments as in the case of the ionization, the absolute value of the recombination for the three bursts is equivalent (the $\mathcal{F}_{rec}$ factor does not change from burst to burst since it is independent on the velocity angle and $v(t_r)$ is fixed by the saddle point equation),
\begin{equation}\label{eq:SameAbsRec}
\frac{| \text{d}_{\ell,m, \pm}^{\text{(rec)}} [\mathbf{v}(t_r^{(2)})] |} { | \text{d}_{\ell,m, \pm}^{\text {(rec)}} [\mathbf{v}(t_r^{(1)})] |}  = \frac{| \text{d}_{\ell,m, \pm}^{\rm (rec)} [\mathbf{v}(t_r^{(3)})] |} { | \text{d}_{\ell,m, \pm}^{\rm (rec)} [\mathbf{v}(t_r^{(1)})] |} = 1.
\end{equation}
The relation between the argument of the recombination of consecutive bursts is
\begin{equation}\label{eq:SameArgRec}
\begin{split}
&\arg \frac{\text{d}_{\ell,m, \pm}^{\rm (rec)} [\mathbf{v}(t_r^{(2)})] } {\text{d}_{\ell,m, \pm}^{\rm (rec)} [\mathbf{v}(t_r^{(1)})] } = - 2(m \pm 1) \pi/3, \\
&\arg \frac{\text{d}_{\ell,m, \pm}^{\rm (rec)} [\mathbf{v}(t_r^{(3)})] }{\text{d}_{\ell,m, \pm}^{\rm (rec)} [\mathbf{v}(t_r^{(1)})]  } = - 4(m \pm 1) \pi/3.
\end{split}
\end{equation}

\section{Derivation of relation between amplitude and phase of different consecutive bursts}
\label{app:amp&phase}
In this section we wish to determine the relation between the amplitudes and phases in Eq.~\eqref{eq:SpectrumIntensity_Definitions} of each of the three bursts in each cycle, in the monochromatic limit.
We will assume that the initial and final state are given by a single orbital, i.e., the sum over $m$ in Eq.~\eqref{eq:FrequencyDipole} runs only over a single index. For a coherent
superposition of two orbitals, like in the case of a $p$ state ($m=1$ and $m=-1$), the results are equivalent if one considers that the electron recombines to the same state
from which it was ionized, i.e., $m_i = m_f$, where $m_i$ and $m_f$ are the initial and final magnetic quantum numbers, respectively. The amplitude and phase of the frequency
dipole in Eq.~\eqref{eq:FrequencyDipole} can be written as
\begin{equation}
\begin{split}
&|\mathbf{D} (N\omega)^{(j)}| = e^{S''^{(j)}-N\omega t_r''^{(j)}} |\Upsilon [\mathbf{v} (t_i^{(j)})]| |\mathbf{d}_{\ell, m} ^{\rm (rec)} [\mathbf{v}(t_r^{(j)})]|, \\
&\arg\left[\text{D}_{\pm} (N\omega)^{(j)}\right] = N\omega t_r'^{(j)} - S'^{(j)} + \\
&+ \arg\{\Upsilon [\mathbf{v} (t_i^{(j)}) \} + \arg\{\text{d}_{\ell,m, \pm} ^{\rm (rec)}[\mathbf{v} (t_r^{(j)})]\}.
\end{split}
\end{equation}
As always, the prime and double prime indicate the real and imaginary component of the corresponding quantity, respectively. In the monochromatic limit, the action is approximately
the same for three consecutive bursts $S^{(1)} \approx S^{(2)} \approx S^{(3)}$.
Due to the symmetry of the field, the real times of ionization and return are shifted by $2\pi/3$ between consecutive bursts, i.e., $t_r'^{(2)} \approx t_r'^{(1)} + \frac{2\pi}{3\omega}$ and $t_r'^{(3)} \approx t_r'^{(1)} + \frac{4\pi}{3\omega}$. 
The imaginary times of ionization and return, on the other hand, do not change for consecutive bursts, i.e., $t_r''^{(1)} \approx t_r''^{(2)} \approx t_r''^{(3)}$. Using the relations \eqref{eq:SameAbsIon}, \eqref{eq:SameArgIon}, \eqref{eq:SameAbsRec}, \eqref{eq:SameArgRec} derived in Appendix \ref{app:IonRec}, we get
\begin{equation}
\begin{split}
&|\mathbf{D} (N\omega)^{(1)}| = |\mathbf{D} (N\omega)^{(2)}| = |\mathbf{D} (N\omega)^{(3)}|, \\
&\arg\left[\text{D}_{\pm} (N\omega)^{(j+1)} \right] - \arg\left[\text{D}_{\pm} (N\omega)^{(j)}\right]  = \frac{2\pi}{3} (N\mp 1),
\end{split}
\end{equation}
as stated in the main text.

\bibliography{biblioContrast}

\end{document}